\documentclass[11pt,a4paper]{article}
\pdfoutput=1
\usepackage{jheppub}

\usepackage[T1]{fontenc}
\usepackage{graphicx}
\usepackage{comment}
\usepackage{grffile}
\usepackage{booktabs}
\usepackage{slashed}
\usepackage{amsmath,amsfonts,bbm,latexsym,amssymb}
\usepackage{epsfig}
\usepackage{array}

\usepackage[utf8]{inputenc}
\usepackage{fancyvrb}
\usepackage{caption}
\usepackage{xcolor}
\usepackage{float}
\usepackage{soul}
\usepackage{subcaption}
\usepackage{orcidlink}
\usepackage{bm}
\usepackage{pifont}

\newcommand{\Z}[1]{\ensuremath{\mathbbm{Z}_{#1}}} % z_N ->\Z{N}

\definecolor{darkgreen}{rgb}{0.13, 0.50, 0.09}

\begin{document}
\begin{flushright}
KA-TP-03-2026 \\
P3H-26-005
\end{flushright}
\vspace*{1cm}

 \title{Reassessing CP Violation in the C2HDM with Machine Learning}

\author[a]{Rafael Boto \orcidlink{0000-0002-9093-205X},}
\author[a]{Karim Elyaouti \orcidlink{0009-0006-0131-5233},}
\author[a]{Duarte Fontes \orcidlink{0000-0002-9358-6500},}
\author[a,b]{Maria Gonçalves \orcidlink{0009-0008-8401-2856},}
\author[a]{Margarete M\"{u}hlleitner \orcidlink{0000-0003-3922-0281},}
\author[c]{Jorge C. Rom\~ao \orcidlink{0000-0002-9683-4055},}
\author[b,d]{Rui Santos \orcidlink{0000-0002-7948-0355},}
\author[c]{and Jo\~{a}o P. Silva \orcidlink{0000-0002-6455-9618}}

\affiliation[a]{Institute for Theoretical Physics,
Karlsruhe Institute of Technology, \\
76128 Karlsruhe, Germany}
\affiliation[b]{Centro de F\'{\i}sica Te\'{o}rica e Computacional, Faculdade de Ci\^{e}ncias, \\
Universidade de Lisboa, Campo Grande, Edif\'{\i}cio C8,
1749-016 Lisboa, Portugal}
\affiliation[c]{Departamento de F\'{\i}sica and CFTP,
Instituto Superior T\'{e}cnico, Universidade de Lisboa, \\
Avenida Rovisco Pais 1, 1049-001 Lisboa, Portugal}
\affiliation[d]{ISEL - Instituto Superior de Engenharia de Lisboa, \\
Instituto Polit\'ecnico de Lisboa 1959-007 Lisboa, Portugal}
\emailAdd{rafael.boto@kit.edu}
\emailAdd{karim.elyaouti@partner.kit.edu}
\emailAdd{duarte.fontes@kit.edu}
\emailAdd{magoncalves@fc.ul.pt}
\emailAdd{margarete.muehlleitner@kit.edu}
\emailAdd{jorge.romao@tecnico.ulisboa.pt}
\emailAdd{rasantos@fc.ul.pt}
\emailAdd{jpsilva@cftp.ist.utl.pt}

\abstract{We provide a study of the parameter space of the complex 2-Higgs Doublet Model (C2HDM), focusing on signs of large CP-violating couplings of the 125~GeV Higgs boson with the fermions. The study is performed utilizing Machine Learning (ML) techniques developed recently for parameter space exploration, including an Evolutionary Strategy Algorithm and Novelty Reward.  We give particular attention to the electron electric dipole moment (eEDM). We confirm that the recently found kite diagrams are  crucial for the outcome of the analysis. Moreover, their use also mitigates the dependence of the results on the scale and scheme choice of the masses in the loop diagrams. We  furthermore point out that, already at the current level of experimental precision, the Barr-Zee diagrams  with charm quark loops must be taken into account. The combined use of kite diagrams and ML techniques allows for the resurrection of large fermion CP-odd couplings  for Type-II and Flipped C2HDM when the 125~GeV Higgs coincides with the second lightest neutral scalar. This arises due to cancellations, typically of the per-mil order, which, moreover, will still be possible for a foreseeable eEDM precision down to $10^{-33}$\, e.cm. For these cases, the constraints on the CP-odd couplings arises from the precision LHC measurements.
}

\makeatother

\maketitle

\section{Introduction}

T.D. Lee showed that a 2-Higgs Doublet Model (2HDM) allows for spontaneous
CP violation in the scalar sector \cite{Lee:1973iz}.
Experiments in $B$-physics 
\cite{BaBar:2001pki,Belle:2001zzw}
guarantee that the CP violation in $W$ boson interactions,
predicted in the Standard Model (SM),
must be large.
This spurred interest in a more general CP-violating framework in the 2HDM.
In order to avoid flavour-changing neutral scalar interactions,
a $\Z2$ symmetry is usually employed \cite{Glashow:1976nt,Paschos:1976ay}.
This leads to a CP-conserving scalar sector in the 2HDM and,
moreover, precludes a decoupling limit.
Both problems may be solved by introducing in the potential a complex
soft-symmetry breaking term ($m_{12}^2$).
This leads to the  complex 2HDM (C2HDM),
which has been extensively explored in the literature;
see, for example, 
Refs.~\cite{Ginzburg:2002wt,Gunion:2002zf,Ginzburg:2004vp,
ElKaffas:2006gdt,Arhrib:2010ju,Barroso:2012wz,Inoue:2014nva,Fontes:2014xva,Grzadkowski:2014ada,ElKaffas:2007rq,Grzadkowski:2009iz,Khater:2003wq,Cheung:2014oaa,Grober:2017gut,Fontes:2015xva, Fontes:2015mea,Abe:2013qla,Chen:2015gaa, Chen:2017com,Fontes:2017zfn,Chun:2019oix,Cheung:2020ugr,Frank:2021pkc,Biekotter:2024ykp, Muhlleitner:2017dkd,Basler:2017uxn,Aoki:2018zgq, Basler:2019iuu,Wang:2019pet,Boto:2020wyf,Altmannshofer:2020shb,Fontes:2021iue, Basler:2021kgq, Abouabid:2021yvw,Fontes:2022izp,Azevedo:2023zkg,Goncalves:2023svb}.

Models with CP violation in the scalar sector must contend with
potentially large contributions to the electron electric dipole moment (eEDM).
The leading contributions arise from two-loop Barr-Zee diagrams \cite{Barr:1990vd} and extensions~\cite{Leigh:1990kf,Gunion:1990ce,Chang:1990sf}. 
In Ref.~\cite{Altmannshofer:2020shb} it was shown that a new class of two-loop diagrams
(dubbed “kite diagrams”) is essential for the 
gauge independence of the final result.
In a subsequent article,
relevant one-loop diagrams were also identified \cite{Altmannshofer:2025nsl}.

The inclusion of all relevant diagrams in the computation of the eEDM is crucial for the following two reasons.
On the one hand, current experiments are extremely accurate: the 90\% confidence-level limit on the eEDM 
reported by the ACME collaboration is
$1.1 \times 10^{-29}\, \textrm{e.cm}$
\cite{ACME:2013pal,ACME:2018yjb,Roussy:2022cmp},
while JILA quotes
$4.1 \times 10^{-30}\, \textrm{e.cm}$ \cite{Roussy:2022cmp}. On the other hand,
in the Feynman gauge, 
each individual diagram can typically yield values of order
$10^{-27}\, \textrm{e.cm}$.
Thus, in order to get substantial large CP violation (e.g. in Higgs-fermion vertices), there needs to be a substantial cancellation among different eEDM diagrams,
often of one part per mil.
%\MMc{Take off the following sentence:}
This forces the consideration of previously overlooked diagrams; for example,
we will show that, given the current eEDM constraints, Barr-Zee diagrams involving virtual charm quarks need to be included.
%are now mandatory. %\MMc{Mandatory for what?}
The expected sensitivity of future eEDM
experiments
has been discussed to be $\mathcal{O}(10^{-33})\,\text{e.cm}$~\cite{Vutha:2018tsz,Ardu:2024bxg,Aiko:2025tbk},
increasing the difficulty of obtaining the necessary cancellations
in a scenario of non-discovery. Using the diagrams included in our study,
we show that the cancellations can still be met for different cases of C2HDM.

The impact of eEDM measurements on the C2HDM parameter space  has been considered 
by a number of authors \cite{Abe:2013qla,Chen:2015gaa,Chen:2017com,Fontes:2017zfn,Chun:2019oix,Cheung:2020ugr,Frank:2021pkc,Biekotter:2024ykp}.
\begin{table}[ht]
  \centering
  \begin{tabular}{|l|c|c|c|c|}\hline
    Type   & I & II& LS& Flipped\\ \hline
$h_1=h_{125}$&$\times$ &$\times$ &$\tau$ & $\underline{\times}$\\\hline
$h_2=h_{125}$ &$\times$ & $\underline{\times}$ &$\tau$ & $\times$\\\hline
$h_3=h_{125}$ &$\times$ &$\times$ &$\tau$ & $\times$\\\hline
  \end{tabular}
  \caption{Results for the
    possibility of large Yukawa couplings from Ref.~\cite{Biekotter:2024ykp}.
A cross means that it is not possible to have
large CP-odd couplings, i.e.~$|c^o| \gtrsim |c^e|$.
The notation $\tau$ means that $c^o/c^e$ is
limited by the direct
searches for CP-violating angular correlations
of tau leptons in
$ h_{125} \rightarrow \tau \bar{\tau}$
decays~\cite{CMS:2021sdq}.
Underlined crosses indicate a change
from allowed ($\checkmark$) to excluded ($\times$)
compared
to the previous analysis carried out
in 2017~\cite{Fontes:2017zfn}.
}
\label{tab:ref}
\end{table}
In Ref.~\cite{Biekotter:2024ykp}, a subset of the authors of this work revisited the C2HDM taking into account the then most recent experimental results in order to quantify the amount of CP-violation that was still possible in the coupling of the SM-like Higgs to the fermions.
In the C2HDM, there are three neutral scalars with mixed CP nature,
denoted by $h_1$, $h_2$, and $h_3$,
ordered by increasing masses.
The Higgs particle found at Large Hadron Collider (LHC) with mass of $125\,\textrm{GeV}$
\cite{ATLAS:2012yve,CMS:2012qbp}
is denoted by $h_{125}$.
All three possibilities ---
$h_1=h_{125}$, $h_2=h_{125}$ and $h_3=h_{125}$ ---
have been considered.
The CP-even (CP-odd) $h_{125}ff$ couplings are denoted by
$c^e_f$ ($c^o_f$),
where $f$ denotes a fermion.
The analysis is performed for each of the four $\Z2$-symmetric C2HDM submodels,
namely, Type-I, Type-II, Lepton-Specific (LS) and Flipped. In Ref.~\cite{Biekotter:2024ykp}, the particular case of the Type-II model with the configuration $h_2 = h_{125}$ received special attention, for reasons detailed below. 
For reference, we display in Tab.~\ref{tab:ref} the results found in \cite{Biekotter:2024ykp} on the possibility of a large CP-odd coupling admixture to the $h_{125}$ Yukawa couplings.
In this work, we revisit this table in light of the new scanning  techniques.
%These Types are denoted in order by $x=1,2,3,4$ in the notation
%``txhy'' to be used below, where $y=1,2,3$ refers to which scalar is taken as $h_{125}$.

Valid points in the C2HDM parameter space were found in
Refs.~\cite{Fontes:2017zfn,Biekotter:2024ykp}
by manually scanning close to the  CP-conserving limit and iteratively
enlarging the pseudoscalar component.
This method was also followed for the $\Z2 \times \Z2$ 3HDM in Ref.~\cite{Boto:2024jgj}.
Using a machine-learning (ML) algorithm with an Evolutionary Strategy and
Novelty Reward mechanism,
it was shown in Ref.~\cite{deSouza:2025bpl} that this method
seriously misses valid regions of the model's parameter space
and the resulting physical implications.
This  motivated us to revisit the C2HDM with
the same ML strategy.
In particular,
we found that the possibility of large CP-odd Higgs-fermion couplings
for the Flipped Type with $h_2=h_{125}$,
which seemed impossible using the usual search techniques,
is revived.

ML has become an essential tool
in particle physics,
driven by the increasing complexity and volume of data from experiments
like those at the LHC. Recent reviews~\cite{Feickert:2021ajf,Shanahan:2022ifi,Plehn:2022ftl}
include the developments in different fronts of Artificial Intelligence in high-energy physics.
We are interested in the validation of a Beyond-the-SM (BSM) theory through
constraints on its parameter space. 
Successful approaches to sampling the parameter space of BSMs have
been developed in cases where training datasets are
available~\cite{Caron:2019xkx,Hollingsworth:2021sii,Goodsell:2022beo,Diaz:2024yfu,AbdusSalam:2024obf,Batra:2025amk,Hammad:2025wst}.
Given the goal set on finding physical  implications that we
do not yet know about in highly constrained scans, we turn to search methods
that do not rely on  an initial set of sampled data.
We consider an Evolutionary Strategy algorithm, first introduced in Ref.~\cite{deSouza:2022uhk} to find valid
and verifiable points of the model,
combined with an anomaly detection used for Novelty
Reward developed in Ref.~\cite{Romao:2024gjx}, to ensure good exploration
of parameter spaces and of the physical observables of the model. 

The work is organized as follows.
In Section~\ref{sec:model},  we recapitulate the symmetries and resulting Lagrangian for the C2HDM, with the possible model choices for the Yukawa couplings. In Section~\ref{sec:eEDM}, we succinctly describe the status of the eEDM calculation in the C2HDM and the experimental bounds. In Section~\ref{sec:sampling}, we complete the list of constraints to be applied to this model, both theoretical and experimental ones, followed by the description of the adopted sampling strategy. Sections~\ref{sec:kite} and \ref{sec:others} are devoted our results, the former focused on the case Type-II with $h_{125}=h_2$, and the latter on other cases. After summarizing our conclusions in Section~\ref{sec:conclusions}, we present the details of on our application of \texttt{HiggsSignals} in Appendix~\ref{app:chisqexclusion}.

\section{The C2HDM}
\label{sec:model}
In this work, we discuss an extension of the SM with a CP-violating scalar sector,
with the addition of a scalar doublet  with a softly broken $\Z2$ symmetry,
known as the C2HDM~\cite{Ginzburg:2002wt,Gunion:2002zf,Barroso:2012wz,Ginzburg:2004vp,ElKaffas:2006gdt,Arhrib:2010ju,Inoue:2014nva,Fontes:2014xva,Grzadkowski:2014ada}.
We follow the description of the model in Ref.~\cite{Fontes:2017zfn} and write the scalar potential as
\begin{eqnarray}
    V &=& m_{11}^2 |\Phi_1|^2 + m_{22}^2 |\Phi_2|^2
- \left(m_{12}^2 \, \Phi_1^\dagger \Phi_2 + h.c.\right)
+ \frac{\lambda_1}{2} (\Phi_1^\dagger \Phi_1)^2 +
\frac{\lambda_2}{2} (\Phi_2^\dagger \Phi_2)^2 \nonumber \\
&& + \lambda_3
(\Phi_1^\dagger \Phi_1) (\Phi_2^\dagger \Phi_2) + \lambda_4
(\Phi_1^\dagger \Phi_2) (\Phi_2^\dagger \Phi_1) +
\left[\frac{\lambda_5}{2} (\Phi_1^\dagger \Phi_2)^2 + h.c.\right] \;,
\end{eqnarray}
where all couplings are real, except for $m_{12}^2$ and $\lambda_5$.
Each of the doublets is written as an expansion around the real vacuum
expectation values (vevs) of the neutral components
$\langle \Phi_i^0 \rangle =v_i / \sqrt{2}\,\,(i=1,2)$,
with $v^2=v_1^2+v_2^2\approx 246~{\rm GeV}$,
\begin{equation}
\Phi_1 = \left(
\begin{array}{c}
\phi_1^+ \\
\frac{v_1 + \rho_1 + i \eta_1}{\sqrt{2}}
\end{array}
\right) \qquad \mbox{and} \qquad
\Phi_2 = \left(
\begin{array}{c}
\phi_2^+ \\
\frac{v_2 + \rho_2 + i \eta_2}{\sqrt{2}}
\end{array}
\right) \;. \label{eq:2hdmdoubletexpansion}
\end{equation}
The minimum conditions can then be written as three equations, 
\begin{eqnarray}
  m_{11}^2 v_1 + \frac{\lambda_1}{2} v_1^3 + \frac{\lambda_{345}}{2} v_1
v_2^2 &=& \textrm{Re} \left(m_{12}^2\right) v_2 \;, \label{eq:mincond1} \\
m_{22}^2 v_2 + \frac{\lambda_2}{2} v_2^3 + \frac{\lambda_{345}}{2} v_1^2
v_2 &=& \textrm{Re} \left(m_{12}^2\right) v_1 \;, \label{eq:mincond2} \\
2\, \mbox{Im} (m_{12}^2) &=& v_1 v_2 \mbox{Im} (\lambda_5)
\;, \label{eq:mincond3}  
\end{eqnarray}
which, for non-zero vevs $v_1$ and $v_2$,
ensure one independent CP-violating phase if 
$\mbox{Im} \left\{\lambda_5^* \left(m_{12}^2\right)^2\right\} \neq 0$
\,\cite{Ginzburg:2002wt,Gunion:2002zf,Barroso:2012wz}.
The diagonalization procedure starts with a
rotation to the Higgs basis~\cite{Georgi:1978ri,Donoghue:1978cj,Lavoura:1994fv,Botella:1994cs},
\begin{equation}
     \left( \begin{array}{c} {\cal H}_1 \\ {\cal H}_2 \end{array} \right) =
 R^T_H \left( \begin{array}{c} \Phi_1 \\
     \Phi_2 \end{array} \right) \equiv
 \left( \begin{array}{cc} c_\beta & s_\beta \\ - s_\beta &
     c_\beta \end{array} \right) \left( \begin{array}{c} \Phi_1 \\
     \Phi_2 \end{array} \right) \;,
\end{equation}
with
\begin{equation}
 \tan \beta \equiv \frac{v_2}{v_1} \;.
\end{equation}
The doublets in the Higgs basis can then be written,
with the Goldstone bosons $G^\pm$ and $G^0$ in ${\cal H}_1$, as
\begin{equation}
 {\cal H}_1 = \left( \begin{array}{c} G^\pm \\ \frac{1}{\sqrt{2}} (v + H^0
     + i G^0) \end{array} \right) \quad \mbox{and} \qquad
 {\cal H}_2 = \left( \begin{array}{c} H^\pm \\ \frac{1}{\sqrt{2}} (R_2
     + i I_2) \end{array} \right) \;.
\end{equation}
The mass matrix for the neutral Higgs states is defined for
$\rho_1$, $\rho_2$ and $\rho_3=I_2$ as,
\begin{equation}
 \;({\cal M}^2)_{ij} = \left\langle \frac{\partial^2 V}{\partial \rho_i
  \partial \rho_j} \right\rangle,
\label{eq:c2hdmmassmat}
\end{equation}
 which can be diagonalized by a general orthogonal matrix $R$~\cite{ElKaffas:2007rq}.
That is,
\begin{equation}
R {\cal M}^2 R^T = \mbox{diag} (m_{1}^2, m_{2}^2, m_{3}^2) \;,
\end{equation}
for which we follow the choice
\begin{equation}
R =
\left(
\begin{array}{ccc}
c_1 c_2 & s_1 c_2 & s_2\\
-(c_1 s_2 s_3 + s_1 c_3) & c_1 c_3 - s_1 s_2 s_3  & c_2 s_3\\
- c_1 s_2 c_3 + s_1 s_3 & -(c_1 s_3 + s_1 s_2 c_3) & c_2 c_3
\end{array}
\right)\, ,
\label{matrixR}
\end{equation}
with the notation $s_i \equiv \sin{\alpha_i}$,
$c_i \equiv \cos{\alpha_i}$ ($i = 1, 2, 3$),
in the ranges
\begin{equation}
- \pi/2 < \alpha_1 \leq \pi/2,
\hspace{5ex}
- \pi/2 < \alpha_2 \leq \pi/2,
\hspace{5ex}
- \pi/2 < \alpha_3 \leq \pi/2.
\label{range_alpha}
\end{equation}
The model contains three neutral particles with no definite CP quantum numbers,
$h_1$, $h_2$ and $h_3$, and two charged scalars $H^\pm$. The masses for the neutral Higgs bosons are ordered such that $m_{1} \le m_{2} \le m_{3}$. The set of independent parameters of the potential sector can be chosen as
\begin{eqnarray}
\tan \beta,\, m_{H^\pm},\,
\alpha_1,\, \alpha_2,\, \alpha_3,\, m_{1},\, m_{2}, \, \textrm{Re}(m_{12}^2).
\end{eqnarray}
With this choice, the mass of the heaviest neutral scalar is a derived quantity,
obeying
\begin{equation}
m_{3}^2 = \frac{m_{1}^2\, R_{13} (R_{12} \tan{\beta} - R_{11})
+ m_{2}^2\ R_{23} (R_{22} \tan{\beta} - R_{21})}{R_{33} (R_{31} - R_{32} \tan{\beta})}.
\label{m3_derived}
\end{equation}
Here, $m_{3}^2$ has to be a positive quantity, which is implemented in the optimization algorithm as a constraint that must be satisfied before the fitting procedure is initiated~\cite{deSouza:2025uxb,deSouza:2025bpl,Boto:2025ovp}.

The imposed $\Z2$ symmetry is motivated by the possibility of forbidding  tree-level flavour-changing neutral couplings known to be very constrained experimentally. The adopted natural flavour conservation mechanism\,\cite{Glashow:1976nt,Paschos:1976ay}  extends the symmetry to the Higgs-fermion Yukawa sector such that each of the three families of fermions couples to one and only one scalar field. Introducing the up-, down- and lepton-type fermion doublets as $\Phi_u$, $\Phi_d$ and $\Phi_\ell$, for  the doublet $\Phi_i\,(i=1,2)$ that respectively couples to up-type, down-type and charged leptons, there are four possible Yukawa types of the softly-broken $\Z2$ symmetric 2HDM:
\begin{itemize}
\item
Type-I:
$\Phi_u=\Phi_d=\Phi_\ell \equiv \Phi_2$
\item
Type-II:
$\Phi_u \equiv \Phi_2 \neq \Phi_d=\Phi_\ell \equiv \Phi_1$
\item
Lepton-Specific (LS)
$\Phi_u=\Phi_d \equiv \Phi_2 \neq \Phi_\ell \equiv \Phi_1$
\item
Flipped
$\Phi_u=\Phi_\ell \equiv \Phi_2 \neq \Phi_d \equiv \Phi_1$.
\end{itemize}
The Yukawa Lagrangian for the neutral scalars can be written as
\begin{equation}
{\cal L}_Y = - \sum_{i=1}^3 \frac{m_f}{v} \bar{\psi}_f \left[ c^e(h_i
  ff) + i c^o(h_i ff) \gamma_5 \right] \psi_f h_i \;, \label{eq:yuklag}
\end{equation}
where $\psi_f$ denote the fermion fields with mass $m_f$. The
coefficients of the CP-even and of the CP-odd part of the Yukawa
coupling, $c^e(h_i ff)$ and $c^o (h_i ff)$, are presented in Tab.~\ref{tab:yukcoup} and will be abbreviated by $c^e_f$ and $c^o_f$, respectively, for the state identified as the $125~{\rm GeV}$ scalar. The different families of fermions are identified by choosing $f$ with the labels $t$, $b$ and $\tau$ for up-type, down-type and leptons, respectively. 
\begin{table}
\begin{center}
\begin{tabular}{rccc} \toprule
& up-type & down-type & leptons \\ \midrule
Type-I & $\frac{R_{i2}}{s_\beta} - i \frac{R_{i3}}{t_\beta} \gamma_5$
& $\frac{R_{i2}}{s_\beta} + i \frac{R_{i3}}{t_\beta} \gamma_5$ &
$\frac{R_{i2}}{s_\beta} + i \frac{R_{i3}}{t_\beta} \gamma_5$ \\
Type-II & $\frac{R_{i2}}{s_\beta} - i \frac{R_{i3}}{t_\beta} \gamma_5$
& $\frac{R_{i1}}{c_\beta} - i t_\beta R_{i3} \gamma_5$ &
$\frac{R_{i1}}{c_\beta} - i t_\beta R_{i3} \gamma_5$ \\
Lepton-Specific & $\frac{R_{i2}}{s_\beta} - i \frac{R_{i3}}{t_\beta} \gamma_5$
& $\frac{R_{i2}}{s_\beta} + i \frac{R_{i3}}{t_\beta} \gamma_5$ &
$\frac{R_{i1}}{c_\beta} - i t_\beta R_{i3} \gamma_5$ \\
Flipped & $\frac{R_{i2}}{s_\beta} - i \frac{R_{i3}}{t_\beta} \gamma_5$
& $\frac{R_{i1}}{c_\beta} - i t_\beta R_{i3} \gamma_5$ &
$\frac{R_{i2}}{s_\beta} + i \frac{R_{i3}}{t_\beta} \gamma_5$ \\ \bottomrule
\end{tabular}
\caption{Yukawa couplings of the Higgs
  bosons $h_i$ in the C2HDM, divided by the corresponding SM Higgs couplings. The expressions correspond to
  $[c^e(h_i ff) +i c^o (h_i ff) \gamma_5]$ from
  Eq.~\eqref{eq:yuklag}. \label{tab:yukcoup}}
\end{center}
\end{table}

The aim of our simulations is to study the CP nature of the $125~{\rm GeV}$ scalar, identified out of the three $h_i$ as $h_{125}$, with a focus on the possibility of large pseudoscalar $c^o_f$ couplings. There are experimental bounds on the CP-odd ($c^o_{t}$) versus CP-even ($c^e_{t}$) 
couplings to the top quark \cite{ATLAS:2020ior}:
\begin{equation}
|\theta_{t}| = |\arctan(c^o_{t} / c^e_{t})| < 43^\circ
\ \ \textrm{at}\ \ 
95\%
\ \ \textrm{CL}.
\label{theta_t}
\end{equation}
In more recent LHC data analyses,
bounds were also placed on the CP-odd ($c^o_{\tau}$) versus CP-even ($c^e_{\tau}$) 
couplings to the tau lepton \cite{CMS:2021sdq,ATLAS:2022akr}:
\begin{equation}
|\theta_{\tau}| = |\arctan(c^o_{\tau} / c^e_{\tau})| < 34^\circ
\ \ \textrm{at}\ \ 
95\%
\ \ \textrm{CL}.
\label{theta_tau}
\end{equation}
There are no direct bounds on the CP-odd coupling to the bottom quark
($c^o_{b}$).
Reference~\cite{Biekotter:2024ykp} found the situation 
summarized in Tab.~\ref{tab:ref}.

The Higgs couplings to the massive gauge bosons $V=W,Z$ are given by
\begin{equation}
    i \, g_{\mu \nu} \, c(h_i V V) \, g_{h^{\textrm{SM}}VV},
\end{equation}
with the SM coupling $g_{h^{\textrm{SM}}VV}$ given by $g\,M_W$ for the $V=W$ case and $g M_Z/\cos{\theta_W}$ for $V=Z$, where $h^{\textrm{SM}}$ denotes the SM Higgs. Since the signs of $c^{e}_f$ and $c^{o}_f$ have no absolute meaning
and are relative to the sign of
$k_V \equiv c(h_{125}VV)$, 
our results are shown with combinations of 
$\text{sign}(k_V) c_f^o$ \textit{vs.}~$\text{sign}(k_V) c_f^e$.

\section{The electron EDM}
\label{sec:eEDM}

% All of the following would be a repetition!:
%The C2HDM has a scalar sector with explicit CP violation.  There is the possibility of a non-zero  eEDM, which is highly constrained experimentally \cite{ACME:2013pal,ACME:2018yjb,Roussy:2022cmp}, with the most recent limit of $4.1\times 10^{-30}~\text{e.cm}$\,\cite{Roussy:2022cmp}. 
% The future sensitivity is discussed to be of order $\mathcal{O}(10^{-33})\,\text{e.cm}$~\cite{Vutha:2018tsz,Ardu:2024bxg,Aiko:2025tbk}.

We now turn to a discussion of the eEDM (the contributions from the muon EDM and from
non-leptonic EDMs are currently less stringent and thus not considered\,\cite{Fontes:2017zfn}). 
For the theoretical predictions, we use the formulae in Refs.~\cite{Barr:1990vd,Yamanaka:2013pfn,Abe:2013qla,Inoue:2014nva,Altmannshofer:2020shb,Altmannshofer:2025nsl}. We stress that the calculation as presented in Refs.~\cite{Altmannshofer:2020shb,Altmannshofer:2025nsl} includes the complete set of non-vanishing contributions to the eEDM in the C2HDM. Although this complete set corresponds to a gauge-independent result, the individual contributions depend in general on the gauge. In this work, after explicitly checking that their sum is gauge independent, we choose the Feynman gauge for these individual contributions. We investigate for the first time the role of the kite diagrams~\cite{Altmannshofer:2020shb} in simulations of the parameter space of the C2HDM satisfying all theoretical and most recent experimental constraints. As discussed above, we also assess the impact of Barr-Zee diagrams with the charm quark (hereafter referred to as charm diagrams). Finally, we comment on diagrams that yield even smaller contributions.

Beyond the question of which diagrams are included, the two-loop eEDM calculation also involves  the choice of the renormalization scheme. In particular, different approaches can be adopted for the renormalization prescriptions of the fermion masses.
The most common choices are  $\overline{\rm MS}$ running
masses at the scale~$M_Z$
($\overline{m}_t(M_Z),
\overline{m}_b(M_Z)$)~\cite{Abe:2013qla,King:2015oxa,Altmannshofer:2020shb}, or pole mass
for the top quark in combination
with the running bottom-quark mass
at the scale $\overline{m}_b$
($m_t,\overline{m}_b(\overline{m}_b)$)~\cite{Fontes:2017zfn,Basler:2017uxn,Basler:2021kgq,Bahl:2022yrs}.%
\footnote{For the tau lepton, we always use the pole mass, $m_{\tau}$. The same choice was made in Ref.~\cite{Biekotter:2024ykp}; however, that reference contains a typo, as it states that two mass schemes were considered for that particle.}
For brevity, in what follows we refer to the first choice as the ``$M_Z$-masses'' scheme and to the second as the ``pole-masses'' scheme.
In Ref.~\cite{Biekotter:2024ykp} it was shown that in the Type-II model with the choice $h_2=h_{125}$, the choice of the renormalisation scheme adopted for the fermion masses in the eEDM calculation is relevant. Using the $M_Z$-masses scheme, it was found that sizable CP-odd components in the
$h_{125} b \bar b$ coupling were incompatible with the combined experimental constraints from the upper limit on the eEDM and from cross-section limits from BSM scalar searches. For the pole-masses scheme, by contrast, there remained the possibility of considerable large pseudoscalar components of the Higgs-fermion couplings. In Section \ref{sec:kite}, we will review this discussion in light of the updated eEDM calculation with kite contributions.

\section{Sampling}
\label{sec:sampling}

\subsection{Theoretical and experimental constraints}
\label{sec:constraints}

With the aim of probing the CP nature of the $125~\mathrm{GeV}$ scalar, we perform an extensive scan of the parameter space subject to both theoretical and experimental constraints. In addition to the  eEDM calculation described in the previous section, the imposed theoretical requirements include boundedness from below of the scalar potential~\cite{Klimenko:1984qx,Ferreira:2009jb,Ivanov:2006yq}, the enforcement of a global electroweak minimum through the discriminant condition that avoids metastable vacua~\cite{Ivanov:2015nea}, and perturbative unitarity~\cite{Kanemura:1993hm,Akeroyd:2000wc,Ginzburg:2003fe}. The effects of new physics on electroweak precision observables are evaluated through the oblique parameters $S$, $T$ and $U$~\cite{Grimus:2007if,Grimus:2008nb}, which are required to agree at the $2\sigma$ level with the global electroweak fit of Ref.~\cite{Baak:2014ora}.

Constraints from Higgs precision measurements are imposed by requiring agreement with the LHC determinations of coupling modifiers~\cite{ATLAS:2019nkf} at $3\sigma$ and signal strengths~\cite{ATLAS:2022vkf} at $2\sigma$, consistent with CMS data~\cite{CMS:2022dwd}, for each individual initial-state $\times$ final-state production channel. 
Direct searches for additional scalar states are taken into account by implementing  the C2HDM in the software framework \texttt{HiggsTools} (HT), version 1.1.3~\cite{Bahl:2022igd}, and running the most recent module \texttt{HiggsBounds} (HB), version 1.7. 
Flavor constraints are incorporated through an explicit calculation of the $b \to s \gamma$ process following Ref.~\cite{Borzumati:1998tg}, requiring agreement with the experimental measurement~\cite{HFLAV:2019otj} at the $3\sigma$ level. As shown in Ref.~\cite{Deschamps:2009rh,Mahmoudi:2009zx,Hermann:2012fc,Misiak:2015xwa,Misiak:2017bgg}, this observable implies the lower bound $m_{H^+} > 580~\mathrm{GeV}$ at $95\%$ CL $(2\sigma)$ in the Type-II C2HDM. The global fit of Ref.~\cite{Haller:2018nnx} confirms this lower bound, and shows that it also holds for the Flipped model. 

Constraints from direct LHC searches for CP violation of the Higgs boson are applied after those from Higgs precision measurements discussed above. Specifically, for the CP-violating couplings of $h_{125}$ to the top quark, we impose the bounds given in Eq.~\eqref{theta_t}~\cite{ATLAS:2020ior}. As for the bounds on the CP-violating couplings of $h_{125}$ to the tau lepton (for short, bounds on $\theta_{\tau}$), we investigate two alternative approaches: either including the likelihood estimate given by the module \texttt{HiggsSignals} (HS) within HT~\cite{Bahl:2022igd}, or considering Eq.~\eqref{theta_tau}~\cite{CMS:2021sdq,ATLAS:2022akr}. 
Concerning the former, a comment is in order. When sampling the parameter space subject to all the constraints described above (with the bounds on $\theta_{\tau}$ implemented via the public version of the HS module), we rapidly encounter a regime in which all points exhibiting sizable pseudoscalar Higgs--fermion couplings are excluded by HS. In App.~\ref{app:chisqexclusion}, we explain in detail the origin of this tension, and how we modified the dataset used by HS to address this issue. In what follows, whenever we consider the HS to impose bounds on $\theta_{\tau}$, we always use the modified dataset. 
A more detailed discussion is provided in App.~\ref{app:chisqexclusion}.
%In these cases, the resulting $\chi^2_{\text{diff}} \equiv \chi^2 - \chi^2_{\mathrm{SM}}$ exceeds 100, indicating a severe tension with the SM reference. We find that this tension is driven predominantly by measurement number~13 in the public HS dataset, corresponding to the experimental result reported in Ref.~\cite{CMS:2021sdq}, which yields a disproportionately large contribution to $\chi^2_{\text{diff}}$. Removing this single measurement eliminates the large $\chi^2_{\text{diff}}$ values, while leaving all other constraints unaffected; in particular, all parameter points with $\chi^2_{\text{diff}} > 100$ in the full dataset remain fully consistent with the Higgs precision measurements included in our analysis (both with the coupling-modifier measurements~\cite{ATLAS:2019nkf} and the signal-strength results~\cite{ATLAS:2022vkf}). For this reason, when we consider the HS to impose bounds on the CP-violating couplings of $h_{125}$ to the tau lepton, we always exclude this specific measurement. A more detailed discussion is provided in App.~\ref{app:chisqexclusion}.

\subsection{Sampling techniques}
\label{sec:sam-tech}

In our results, we shall be mainly interested in the Type-II and Flipped models. For a generic assignment of $h_i$ to $h_{125}$, the sampling regions can be presented in a compact way for both models.
\begin{equation}
\begin{array}{lcllcl}
G_F&=& 1.16638 \,\times  10^{-5} \,\mbox{GeV}^{-2} & \quad M_Z &=& 91.1876 \mbox{ GeV} \\[0.1cm]
\alpha_{\text{em}}^{-1} (M_Z) &=& 127.92 & \quad M_W &=& 80.385 \mbox { GeV}\\[0.1cm]
\alpha_{\text{em}}^{-1}   &=& 137.035999 & \quad m_\tau &=& 1.77682  \mbox{ GeV}\\[0.1cm]
m_t &=& 172.5 \mbox{ GeV} & \quad \overline{m}_t(M_Z) &=& 163 \mbox{ GeV} \\[0.1cm]
\overline{m}_b(\overline{m}_b) &=& 4.7800 \mbox { GeV} &\quad \overline{m}_b(M_Z) &=& 2.88 \mbox{ GeV} \\[0.1cm]
\Gamma_W &=&  2.08430 \mbox { GeV} &\quad \Gamma_Z &=& 2.49427 \mbox{ GeV} 
\end{array}
\end{equation}
After setting $m_{{125}} = 125.09~{\rm GeV}$, we can write the sampling of the independent parameters as 
\begin{equation}
\begin{split}
&m_{1<h_{125}} \in [15.0,122.5]\, \textrm{GeV}\,; 
\quad m_{2>h_{125}} \in [127.5,1000]\, \textrm{GeV}\,; 
\quad m_{H^\pm} \in [580,1000] \,\textrm{GeV}\,; 
\\*[3mm]
&\tan{\beta} \in [0.3,30.0]\,;
\quad \textrm{Re}(m_{12}^2)
\in [\pm 10^{-1},\pm 10^7]\, ; \quad \alpha_{1}, \alpha_{2},
\alpha_{3}
\in \left[-\frac{\pi}{2},\frac{\pi}{2}\right]\,. 
\end{split}
\label{scaned_region}
\end{equation} 
However, a purely random sampling of the parameter space is highly inefficient in identifying configurations with large pseudoscalar fermion-Higgs couplings. As discussed above, the mass parameter $m_{3}^2$ is a derived quantity and it can become negative in parts of the sampled space. These points must be discarded as they are unphysical. In addition, for the choice $h_2 \equiv h_{125}$ the heavy neutral scalar is further required to be approximately degenerate in mass with the charged Higgs in order to comply with experimental constraints~\cite{Haber:2010bw}. Still, obtaining an experimentally viable solution in the resulting reduced parameter region is a slow process.

We therefore adopt two complementary strategies, which we label Strategy~1 and Strategy~2 in the following. In the first, we perform a guided scan in which the parameter space is explored close to the phenomenological boundary, iteratively increasing the pseudoscalar component of the Higgs-fermion couplings~\cite{Fontes:2017zfn,Biekotter:2024ykp,Boto:2024jgj}. In the second, we employ an AI-based black-box optimisation technique, originally introduced in Ref.~\cite{deSouza:2022uhk}, subsequently applied to a real 3HDM in Ref.~\cite{Romao:2024gjx}, and later extended to 3HDMs with explicit CP violation in the scalar sector~\cite{deSouza:2025bpl,Boto:2025ovp}.

In Strategy~1, the sampling is initiated from points for which the squared mass of the third neutral scalar is already physical. In contrast, the ML method of Strategy~2 --- which does not rely on pre-existing training data --- implements a hierarchical optimization procedure in which the positivity of the squared mass is enforced as a hard constraint within the loss function. Only after this requirement is satisfied does the algorithm proceed to optimize the remaining theoretical and experimental constraints~\cite{deSouza:2025uxb,Boto:2025ovp,deSouza:2025bpl}.

The sampling algorithm employed in Strategy~2 is the Covariant Matrix Adaptation Evolutionary Strategy (CMA-ES)~\cite{Hansen2001,hansen2023cmaevolutionstrategytutorial}. By construction, individual CMA-ES runs are statistically independent: each run is initialized with a  new  choice of the mean vector and covariance matrix, and the optimization proceeds using only the information generated within that run. While this fully blind initialization ensures independence, it was found to lead to an uneven coverage of the parameter space, with distinct regions being explored in isolation and the overall sampling density remaining sparse in intermediate domains.
To mitigate this limitation and improve the continuity and efficiency of the exploration, CMA-ES also allows for initialization using the mean vector and covariance matrix obtained from previous optimization runs, hereafter referred to as \emph{seeded runs}. When combined with an integrated anomaly-detection--based novelty reward mechanism, implemented via a Histogram-Based Outlier Score (HBOS)~\cite{HBOS} evaluated on a selected subset of parameters or observables of interest, this strategy yields a more uniform and coherent exploration of the model’s parameter space and its associated physical implications.

We conclude this section with two final remarks.
First, all datasets used to produce the plots in this paper --- under both Strategy~1 and Strategy~2 --- are generated by imposing all constraints described in Sec.~\ref{sec:constraints}, with the exception of those associated with $\theta_{\tau}$. As discussed at the end of Sec.~\ref{sec:constraints}, two alternative implementations of the $\theta_{\tau}$ constraints are considered; hence, whenever these constraints are imposed, we explicitly state which alternative is adopted.
Second, all plots presented in this work that rely on Strategy~2 are based on samples containing at least $\mathcal{O}(10^7)$ points.

\section{\texorpdfstring{Results for Type-II with $h_2=h_{125}$}{Results for Type-II with h2=h125}}
\label{sec:kite}

Motivated by the discussion above on the choice of the fermion-mass renormalization schemes in the eEDM calculation, we focus in this section on the Type-II model with $h_2 \equiv h_{125}$. 
Besides the usual Barr-Zee diagrams \cite{Barr:1990vd,Leigh:1990kf,Gunion:1990ce,Chang:1990sf} (whose diagrams with fermion loops include only third generation fermions), we start by including only the kite diagrams in our eEDM calculation. This study is carried out in Secs.~\ref{sec:Strategy1} and~\ref{sec:Strategy2}, which implement Strategy~1 and Strategy~2, respectively.
At the end of Sec.~\ref{sec:Strategy2}, we discuss the role of the contributions of other diagrams to the eEDM.
In Sec.~\ref{sec:smallmasses}, we then turn to a specific region of the parameter space characterized by small values of the scalar mass $m_1$. Finally, Sec.~\ref{sec:charm} is devoted to a detailed investigation of the role played by the charm diagrams in the eEDM calculation.

\subsection{Phenomenology with Strategy~1}
\label{sec:Strategy1}

\begin{figure}[h!]\centering
\begin{subfigure}[b]{0.45\linewidth}
\centering
\includegraphics[width=1.00\textwidth]{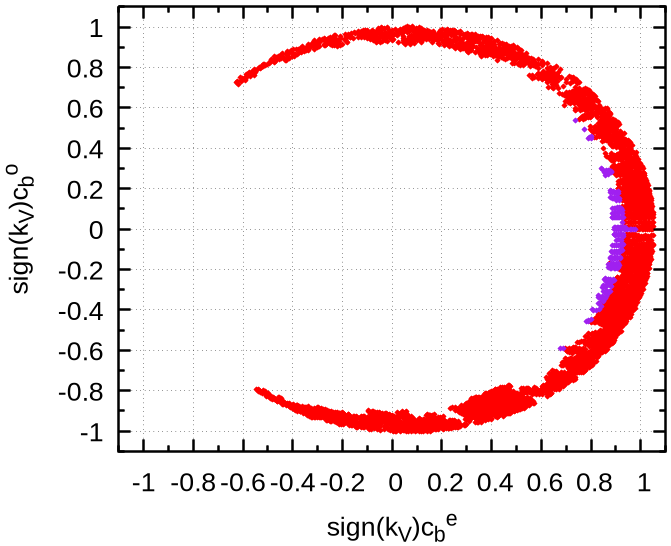}
\end{subfigure}
\hspace{7mm}
\begin{subfigure}[b]{0.45\linewidth}
\centering
\includegraphics[width=1.00\textwidth]{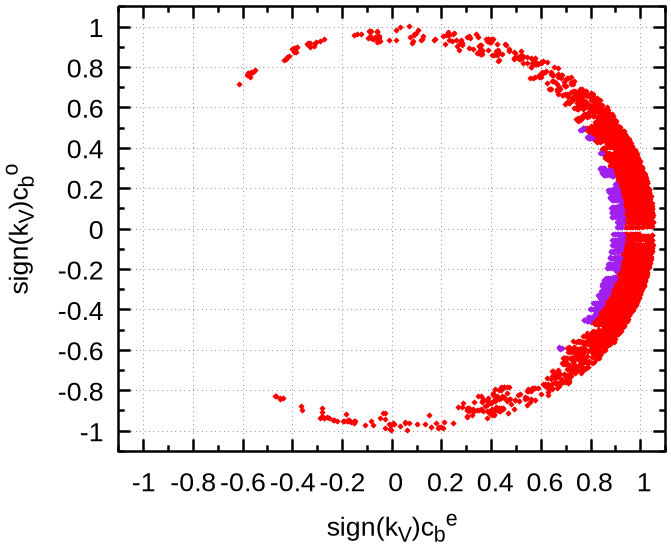}
\end{subfigure}
\caption{
Points on the plane $\text{sign}(k_V)c_b^o \,\, \text{vs.} \,\, \text{sign}(k_V)c_b^e$ obtained with Strategy~1 in the Type-II C2HDM with $h_{125}=h_2$. The purple points include all constraints, including those relative to $\theta_\tau$ implemented via HS, while the red points include all constraints except those relative to $\theta_\tau$. Left panel: current eEDM limit $4.1\times 10^{-30} \,\mathrm{e\cdot cm}$. Right panel: projected eEDM limit $1.0\times 10^{-33} \,\mathrm{e\cdot cm}$. See text for details.}
\label{fig:edmvary}
\end{figure}

Figure~\ref{fig:edmvary} shows two plots of the plane $\text{sign}(k_V)c_b^o \,\, \text{vs.} \,\, \text{sign}(k_V)c_b^e$ generated with Strategy~1,  imposing the current eEDM limit (left) and imposing a projected eEDM limit $1.0\times 10^{-33} \,\mathrm{e\cdot cm}$~\cite{Vutha:2018tsz,Ardu:2024bxg,Aiko:2025tbk} (right).
In the calculation of the eEDM in both panels, the $M_Z$ masses are used and, as discussed above, only the usual Barr-Zee diagrams and the kite diagrams are included. %\MM{and charm ones are not $\to$ while neglecting diagrams with charm quarks}
The red points correspond to parameter points that fulfill all constraints discussed in section \ref{sec:sampling} except those on $\theta_{\tau}$, while the purple points additionally impose those constraints using HS. 
The figure shows that, for both eEDM scenarios considered in the panels, the dominant limitation on the magnitude of the pseudoscalar component of $h_{125}$ arises from the bounds on $\theta_{\tau}$. The figure also shows that the valid (purple) points are all close to the SM solution $\big(\text{sign}(k_V)c_b^e, \text{sign}(k_V)c_b^o\big)=(1,0)$; in particular, no points can be found close to the so-called ``wrong-sign'' solution $\big(\text{sign}(k_V)c_b^e, \text{sign}(k_V)c_b^o\big)=(-1,0)$. 
Finally, we confirm that the points sampled are in agreement with the public code ScannerS~\cite{Muhlleitner:2020wwk}, now also updated to include the kite diagrams in the eEDM calculation.

\begin{figure}[h!]\centering
\begin{subfigure}[b]{0.47\linewidth}
\centering
\includegraphics[width=1.00\textwidth]{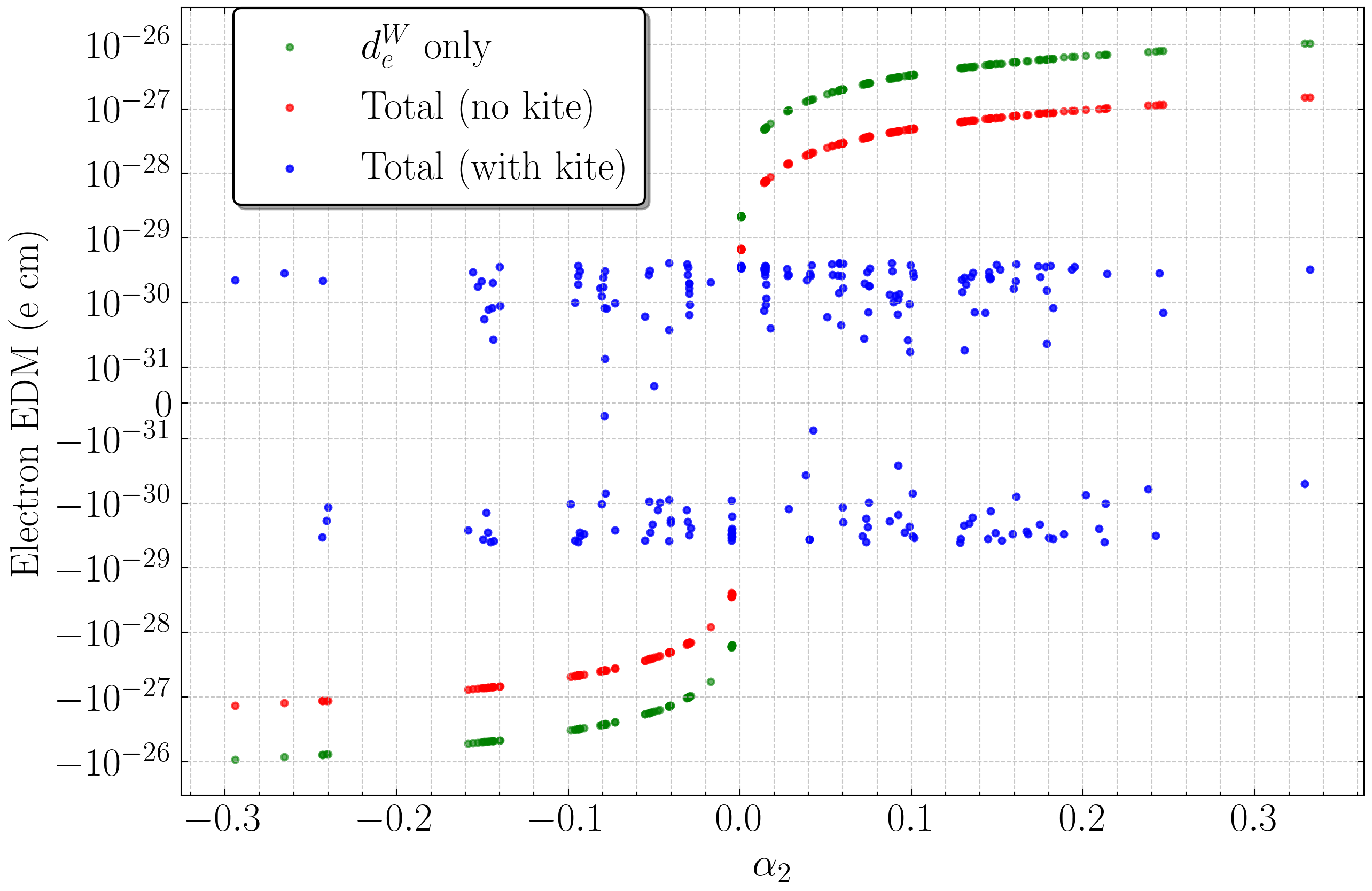}
\end{subfigure}
\hspace{3mm}
\begin{subfigure}[b]{0.47\linewidth}
\centering
\includegraphics[width=1.00\textwidth]{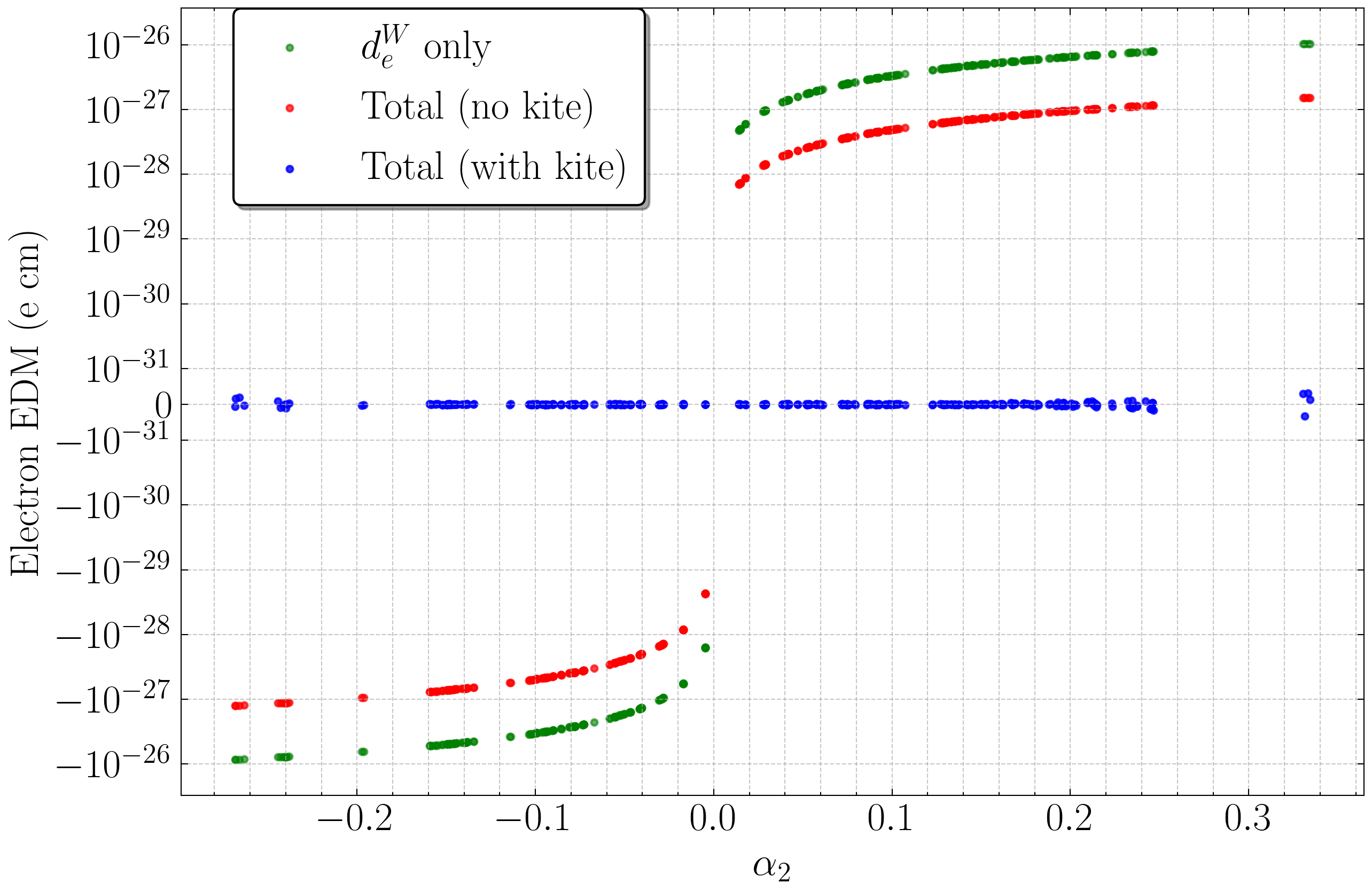}
\end{subfigure}
\caption{Different contributions to the eEDM for the set of points sampled: total contribution (blue), total contribution except for the kite diagrams (red), contribution of the dominant class of diagrams (green). Left panel: current eEDM limit $4.1\times 10^{-30} \,\mathrm{e\cdot cm}$. Right panel: projected eEDM limit $1.0\times 10^{-33} \,\mathrm{e\cdot cm}$ .
}
\label{fig:edmvary_kite}
\end{figure}

In Fig.~\ref{fig:edmvary_kite}, we assess the impact of the kite diagrams on the eEDM prediction. The blue points show the total predicted value of the eEDM, the red ones correspond to the same prediction with the kite-diagram effects subtracted, and the green points represent the class of diagrams yielding the largest contribution to the eEDM for each sampled parameter point. From this comparison, we confirm the result of Ref.~\cite{Altmannshofer:2020shb} that the inclusion of the kite diagrams induces $\mathcal{O}(1)$ corrections to the eEDM prediction for the dataset sampled.

\subsection{Phenomenology with Strategy~2}
\label{sec:Strategy2}

As mentioned in the Introduction,
standard scans can completely miss certain valid regions of the parameter space --- regions where new phenomenology may be possible. Having discussed in the previous section the values found on the plane $\text{sign}(k_V)c_b^o \,\, \text{vs.} \,\, \text{sign}(k_V)c_b^e$ with the standard scans of Strategy~1,
we now turn to ML techniques. The datasets considered below include optimization runs seeded with a subset of points obtained from earlier unseeded runs. The simulations are designed to determine the maximal allowed size of the CP-odd fermionic coupling, imposing the bounds on $\theta_{\tau}$ either through HS or through Eq.~\eqref{theta_tau}. All optimization runs incorporate a novelty-reward mechanism targeting large CP-odd fermion couplings.

\begin{figure}[h!]\centering
\begin{subfigure}[b]{0.47\linewidth}
\centering
\includegraphics[width=1.0\textwidth]{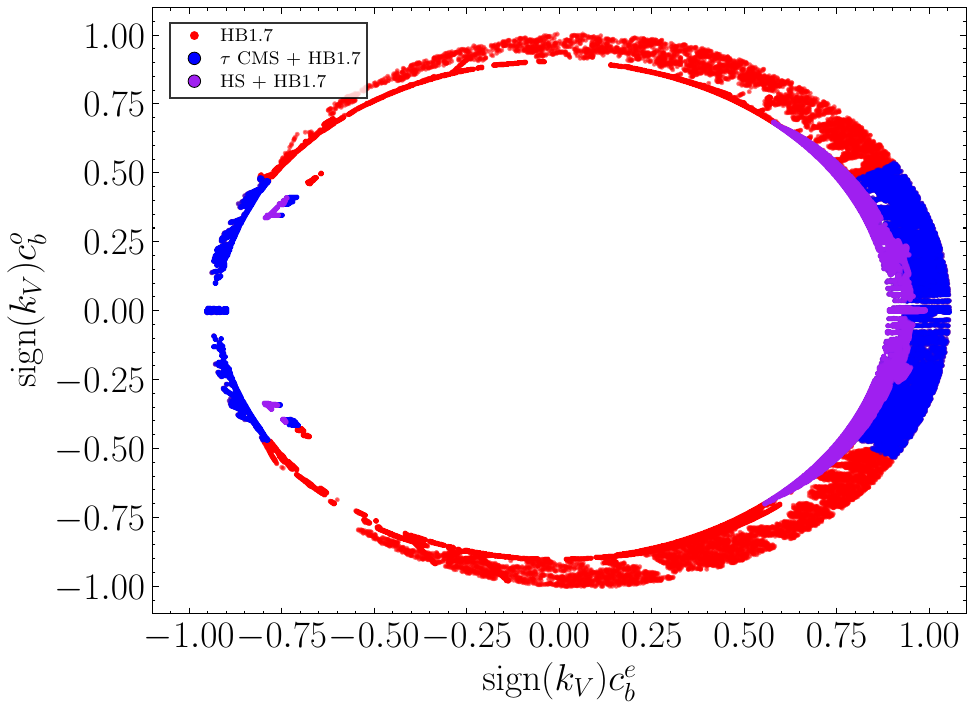}
\end{subfigure}
\hspace{4mm}
\begin{subfigure}[b]{0.47\linewidth}
\centering
\includegraphics[width=1.0\textwidth]{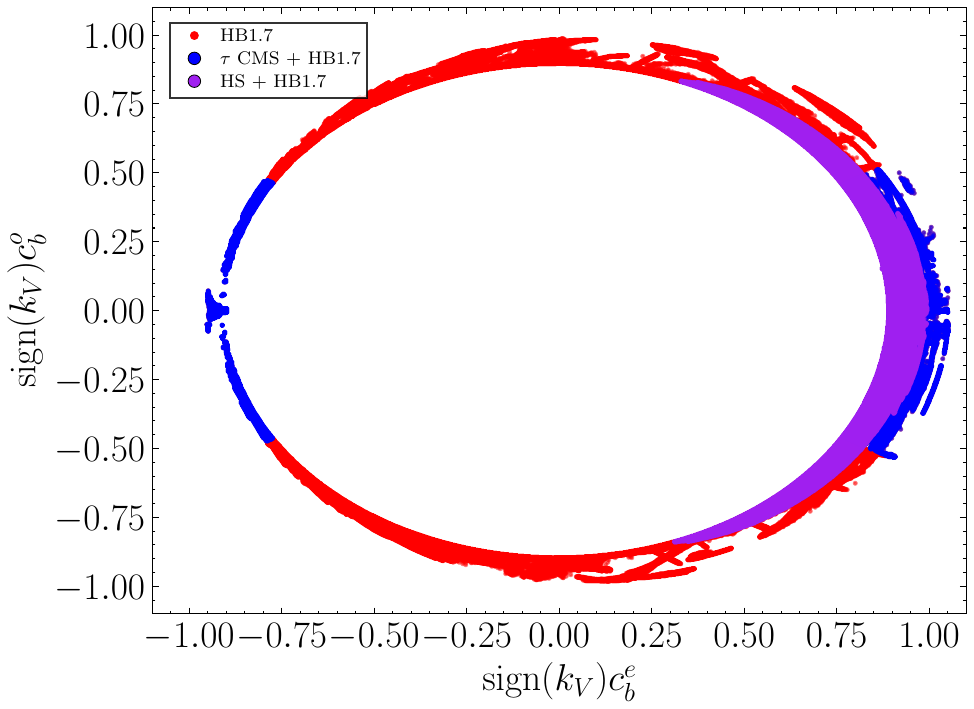}
\end{subfigure}
\caption{Points on the plane $\text{sign}(k_V)c_b^o \,\, \text{vs.} \,\, \text{sign}(k_V)c_b^e$ obtained with Strategy~2 in the Type-II C2HDM with $h_{125}=h_2$. Red points pass all constraints except those on $\theta_\tau$; the remaining points additionally impose the latter constraint: either via HS (in purple) or via Eq.~(\ref{theta_tau}) (in blue). Left: eEDM calculated in the $M_Z$-masses scheme. Right: eEDM calculated in the pole-masses scheme.}
\label{fig:MLpoints_17_tau}
\end{figure}

The results are shown in Fig.~\ref{fig:MLpoints_17_tau}. As before, the calculation of the eEDM is performed including the usual Barr-Zee diagrams and the kite diagrams. We now differentiate the eEDM calculation performed with  the $M_Z$-masses scheme (left panel) from the calculation performed with the pole-masses scheme (right panel).\footnote{For the definition of the $M_Z$-masses and the pole-masses scheme, cf.~Sec.~\ref{sec:eEDM}.}
Points shown in red satisfy all constraints described in Sec.~\ref{sec:sampling}, with the exception of those on $\theta_{\tau}$.
The remaining points additionally impose the latter bounds: either via HS (in purple) or via Eq.~(\ref{theta_tau}) (in blue). This figure should be directly compared with Fig.~\ref{fig:edmvary}. Several noteworthy features emerge from this comparison.

First, Fig.~\ref{fig:MLpoints_17_tau} clearly demonstrates the substantial advantage of the ML-based search over the standard scanning strategy of Fig.~\ref{fig:edmvary}. The ML approach uncovers a significantly larger region of parameter space with large CP-odd couplings. In particular, it reveals a viable region in the vicinity of the wrong-sign solution --- which, as noted above, is entirely absent in the standard scan.

Second, the qualitative features of the ML results are similar for both $M_Z$-masses and pole-masses schemes. Dedicated tests show that this similarity disappears once the kite diagrams are excluded, in agreement with the findings of Ref.~\cite{Biekotter:2024ykp} (see, in particular, the comparison between the right panel of Fig.~1 and the left panel of Fig.~2 therein). This provides further confirmation of the central role of the kite diagrams, as previously emphasized in Ref.~\cite{Altmannshofer:2020shb}. Their importance is not unexpected, given that, as discussed before, these diagrams are required to ensure gauge invariance of the eEDM calculation~\cite{Altmannshofer:2020shb}. As a consequence, non-negligible values of $|c_b^o|$ in the Type-II C2HDM with $h_{125} \equiv h_2$ are found to be phenomenologically viable, constrained primarily by the measurements of $\theta_\tau$.

Third, Fig.~\ref{fig:MLpoints_17_tau} illustrates the impact of the two different implementations of the $\theta_\tau$ constraint. The blue points cover the same width of the ellipse as the red points, but they stop abruptly for $|\theta_{\tau}| \ge 34^\circ$ due to the constraint of Eq.~(\ref{theta_tau}). The region covered by the purple points, by contrast, has a narrower width than the red points, and they are not constrained by the abrupt cut at $|\theta_{\tau}| = 34^\circ$. Both these features result from the likelihood estimate given by HS.

Finally, in the $M_Z$-masses scheme shown in the left panel of Fig.~\ref{fig:MLpoints_17_tau}, we observe isolated regions near $|c_b^o| \sim 0.4$ and $\text{sign}(k_V)c_b^e \sim -0.75$ that are absent in the pole-masses scheme displayed in the right panel. These points correspond to comparatively light scalar masses, $m_1 \sim 50~\mathrm{GeV}$. We return to a detailed discussion of this region in Sec.~\ref{sec:smallmasses} below.

\begin{figure}[h!]\centering
\includegraphics[width=0.5\textwidth]{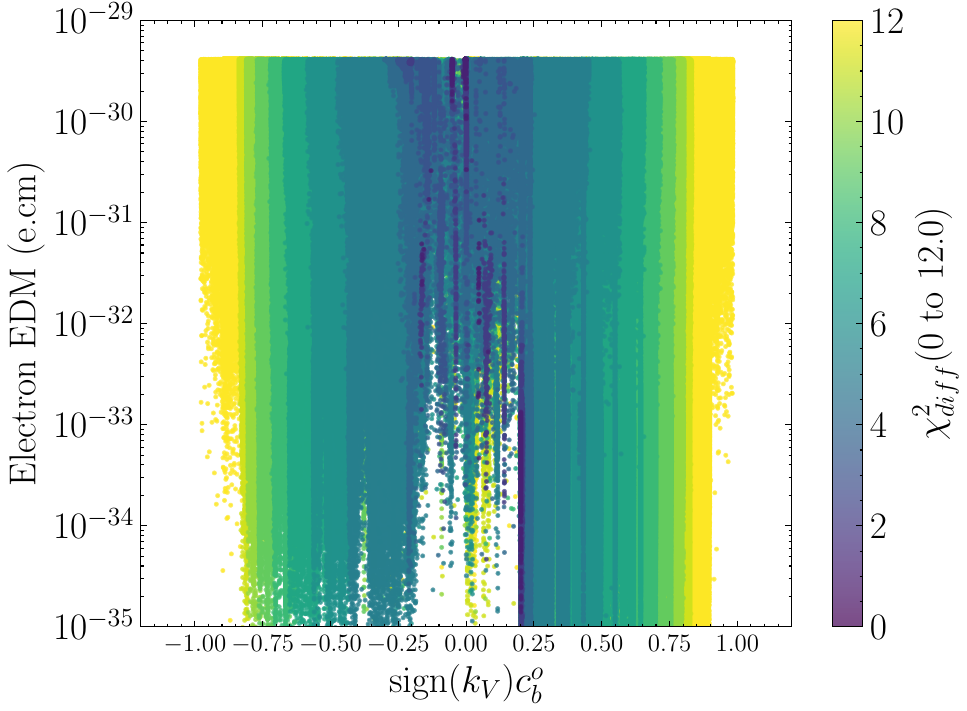}
\caption{Points obtained with Strategy~2 in the Type-II C2HDM with $h_{125}=h_2$, on the plane eEDM prediction (calculated with the pole-masses scheme) vs. $\text{sign}(k_V) c_b^o$. The color code shows the lowest possible $\chi^2$
calculated with HS.}
\label{fig:MLpoints-edm}
\end{figure}

An interesting question concerns the possibility of accommodating increasingly stringent eEDM bounds while still obtaining parameter points with the largest $|c_b^o|$ components compatible with current experimental constraints. This issue is illustrated in Fig.~\ref{fig:MLpoints-edm}.%
\footnote{As before, the calculation of the eEDM is performed including the usual Barr-Zee diagrams and the kite diagrams, assuming the pole-masses scheme. Furthermore, and as usual, the plot is generated from datasets satisfying all constraints described in Sec.~\ref{sec:constraints} except those on $\theta_{\tau}$. On the other hand, the figure does not commit to a specific prescription for imposing the $\theta_{\tau}$ constraints.
If one adopts the alternative based on Eq.~\eqref{theta_tau}, only the points lying within the range of $\text{sign}(k_V) c_b^o$ spanned by the blue points in the right panel of Fig.~\ref{fig:MLpoints_17_tau} are considered valid. If, instead, the HS alternative is followed, the color coding of the figure, representing the $\chi^2_{\text{diff}}$ obtained from HS, allows the $\theta_{\tau}$ constraint to be imposed at the desired confidence level within that framework.} 
The plot shows that,
irrespective of how stringent the eEDM bound becomes --- even down to $10^{-35}\,\mathrm{e\!\cdot\!cm}$ --- it is always possible to identify regions of parameter space that allow for sizable $|c_b^o|$ components, provided the parameters are tuned so that large cancellations conspire to satisfy progressively stronger eEDM constraints.

Although not shown explicitly, we also  performed dedicated runs that led to further observations. First, percent-level variations in the input mass values still permit the identification of parameter sets for which the required cancellations to satisfy the eEDM bounds occur. 
It is found, however, that such small changes require significant  changes in the remaining input parameters in order to still fulfill the required cancellations. This behavior is not unexpected, given that the cancellations involve contributions from very different diagrams. Second, once the eEDM sensitivity reaches the level of $10^{-33}\,\mathrm{e\!\cdot\!cm}$, the contributions of several types of diagrams not included so far in our eEDM calculation (such as Barr-Zee diagrams with quarks beyond the third generation, as well as one-loop diagrams) become numerically relevant.

\subsection{\texorpdfstring{Isolated regions with small $h_{1}$ masses}{Isolated regions with small h1 masses}}
\label{sec:smallmasses}

\begin{figure}[h!]\centering
\includegraphics[width=0.5\textwidth]{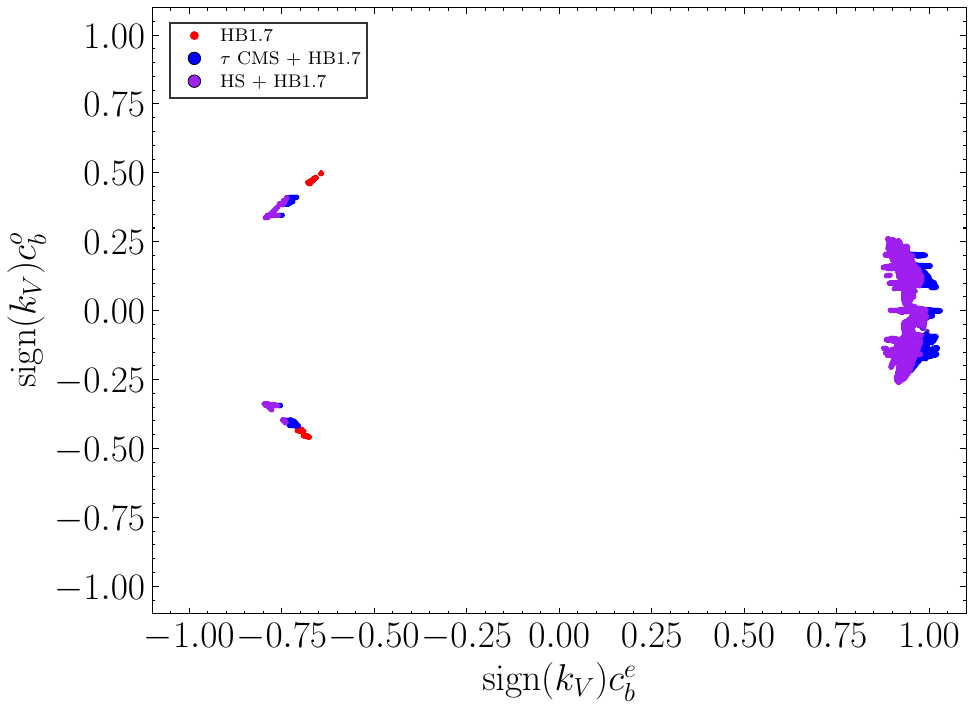}
\caption{The same as the left panel of Fig.~\ref{fig:MLpoints_17_tau}, but considering
exclusively points with $m_{1} \leq 60\textrm{GeV}$.
}
\label{fig:lowMH1}
\end{figure}

As discussed above, the isolated regions located around
$|c^o_b| \sim 0.4$ and $\text{sign}(k_V) c^e_b \sim -0.75$ in the plot with the $M_Z$-masses scheme of Fig.~\ref{fig:MLpoints_17_tau} (left panel) correspond to scenarios with small scalar masses, $m_1 \sim 50\,\mathrm{GeV}$. We probe this feature in more detail in Fig.~\ref{fig:lowMH1}, which displays the subset of points from the left panel of Fig.~\ref{fig:MLpoints_17_tau} with $m_1 < 60\,\mathrm{GeV}$.
The figure reveals a region close to the SM-like point
$\big(\text{sign}(k_V) c_b^e, \text{sign}(k_V) c_b^o\big) = (1,0)$,
as well as an isolated region near $|c^o_b| \sim 0.4$ and $\text{sign}(k_V)  c^e_b \sim -0.75$. This latter region is disconnected from the vicinity of the wrong-sign solution $\big(\text{sign}(k_V) c_b^e, \text{sign}(k_V) c_b^o\big) = (-1,0)$, for which no parameter points with small $m_1$ masses survive. In fact, the isolated region corresponds to a very narrow mass window, $51\,\mathrm{GeV} \lesssim m_1 \lesssim 53\,\mathrm{GeV}$.

We note that the analogous plot obtained using the pole-masses scheme does not exhibit these isolated regions. This behavior is easy to understand. As discussed above, the parameter regions compatible with the eEDM constraints arise from very precise cancellations among different diagrammatic contributions, typically at the per-mil level. Thus, small variations in the input mass parameters lead to large shifts in the remaining parameter space, making it unsurprising that a narrow region present for the $M_Z$-masses scheme disappears for the pole-masses scheme.

\begin{figure}[h!]\centering
\begin{subfigure}[b]{0.47\linewidth}
\centering
\includegraphics[width=1\textwidth]{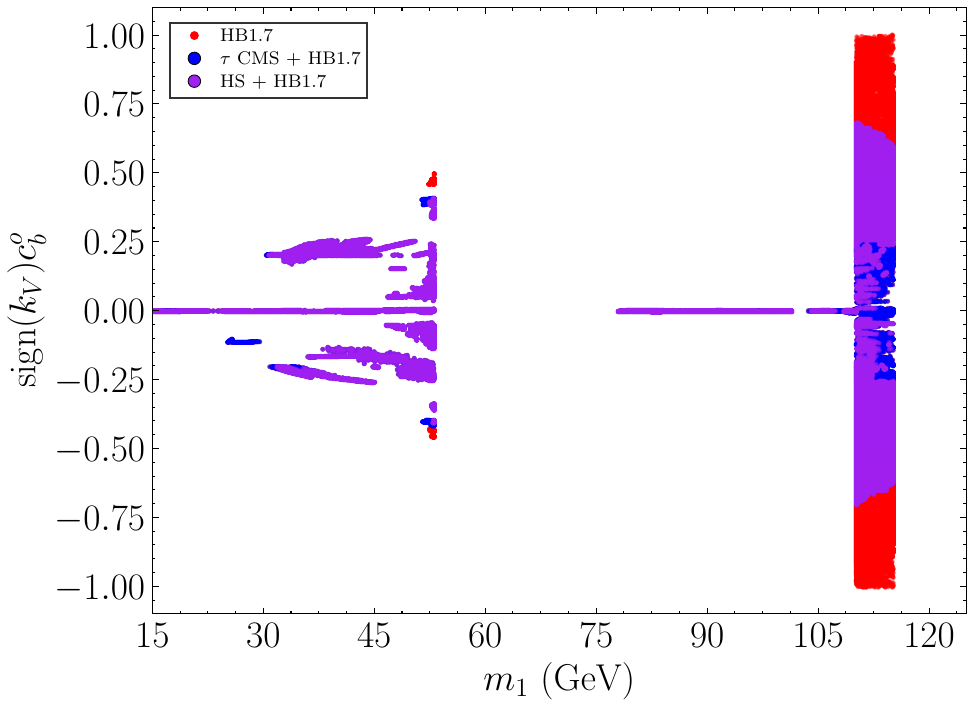}
\end{subfigure}
\hspace{2mm}
\begin{subfigure}[b]{0.47\linewidth}
\centering
\includegraphics[width=1\textwidth]{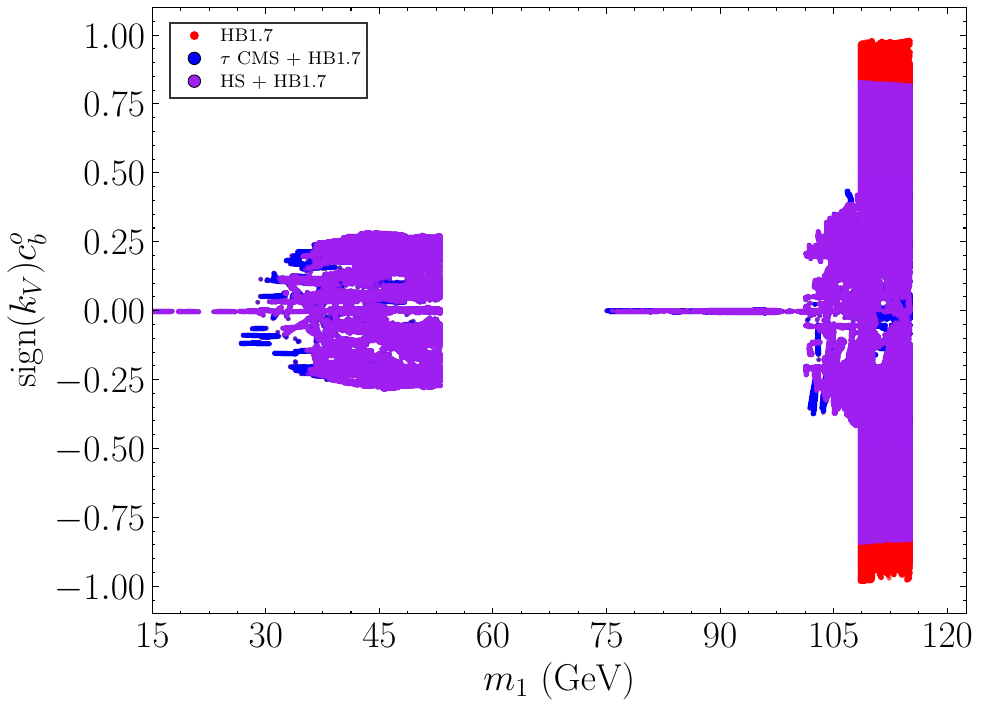}
\end{subfigure}
\caption{The same as in Fig.~\ref{fig:MLpoints_17_tau}, but for the plane $\text{sign}(k_V) c_b^o$ vs.~$m_1$.}
\label{fig:MLpoints_17_massh1}
\end{figure}

To investigate the range of values taken by $m_1$, Fig.~\ref{fig:MLpoints_17_massh1} displays all points shown in Fig.~\ref{fig:MLpoints_17_tau}, now projected onto the plane $\operatorname{sign}(k_V)\, c_b^o$ versus $m_1$. Intermediate values of $m_1$ are excluded by HB through direct searches for additional Higgs bosons, as reported in Refs.~\cite{CMS:2018rmh,CMS:2022goy,ATLAS:2022enb}.
From the left panel, one observes side bands with $|c_b^o| > 0.3$ clustered around $m_1 \sim 50\,\mathrm{GeV}$, corresponding to the isolated regions previously identified in Figs.~\ref{fig:MLpoints_17_tau}~(left) and~\ref{fig:lowMH1}. These regions lie close to the current exclusion boundary. Given the strong sensitivity of these features to the precise input parameters, we do not attribute particular physical significance to the isolated regions appearing when the fermion masses are defined in the $M_Z$-masses scheme.

%\begin{figure}[htb]
%  \centering
%    \includegraphics[width=0.5\textwidth]{Figures/co-mh1-T2-h2-NoKite-MZ.pdf}
%  \caption{noML's points on no kite - MZ. 
%}
%  \label{fig:co-mh1-T2-h2-NoKite-MZ}
%\end{figure}
%
%Did a simulation without HB and it was possible to get the
%intermediate mass region.
%All points do not pass HB when checked,
%with the exclusion occurred due to ATLAS:2022enb for both mass scale choices.

\subsection{The charm diagrams}
\label{sec:charm}

Up to now, we included in our calculations for the eEDM the new kite diagrams from Ref.~\cite{Altmannshofer:2020shb}, besides the usual  Barr-Zee diagrams (which, we recall, contain only fermions of the third generation). One could wonder whether a Barr-Zee diagram with an intermediate charm quark would be important given the current experimental bounds. As we will now show, this is indeed the case, despite their mass suppression.

\begin{figure}[h!]\centering
\includegraphics[width=0.55\textwidth]{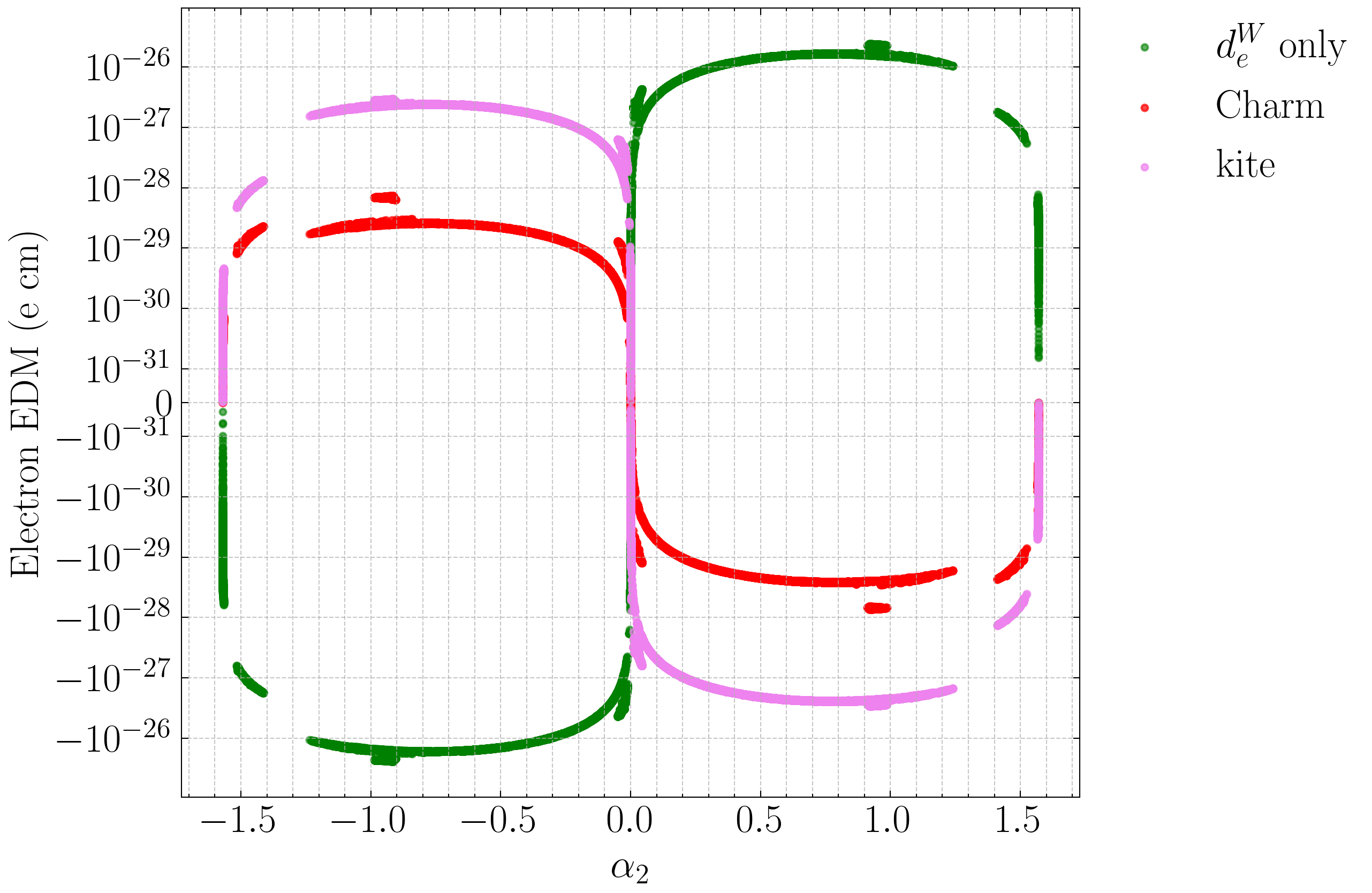}
\vspace*{3mm}
\caption{Points of the left panel of Fig.~\ref{fig:MLpoints_17_tau}, now projected on the plane eEDM vs.~$\alpha_2$. Besides the contributions from the kite diagrams (pink) and the dominant diagrams (green), both of which were included in the calculation of the eEDM in Fig.~\ref{fig:MLpoints_17_tau}, the plot also shows the contribution of the charm diagrams (red). 
}
\label{fig:MLpoints-edm-MZ}
\end{figure}

We have taken some of the points in the left panel of Fig.~\ref{fig:MLpoints_17_tau}, which  obey the current eEDM bounds and which were calculated \textit{without} taking into consideration the charm diagrams; we then calculated, for those points, what the contribution of the charm diagrams would be. For the charm quark, we chose the running $\overline{\mbox{MS}}$ mass $\overline{m}_c$ at the scale $\overline{m}_c$, namely $\overline{m}_c(\overline{m}_c)=1.2$~GeV \cite{Peset:2018ria}. The result is shown in Fig.~\ref{fig:MLpoints-edm-MZ}, which displays separately the contributions from the dominant class of diagrams (green), the kite diagrams (pink) and the charm diagrams (red).
As was already discussed in the context of Fig.~\ref{fig:edmvary_kite}, it is clear that the kite diagrams are crucial for the cancellation.
The plot also shows that the charm contribution comes in at the
order of a few times $10^{-29}\,\text{e.cm}$, well above the current experimental level of $4.1 \times 10^{-30}\, \textrm{e.cm}$ \cite{Roussy:2022cmp}. Thus, it spoils the cancellation,
and the points become disallowed.

\begin{figure}[h!]\centering
\begin{subfigure}[b]{0.47\linewidth}
\centering
\includegraphics[width=1.0\textwidth]{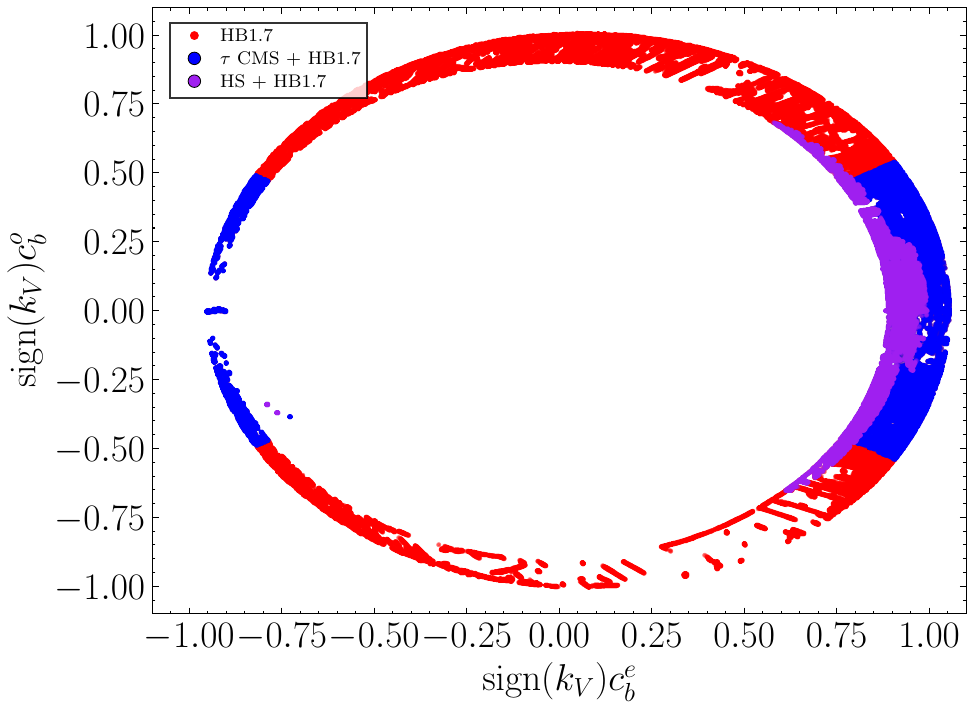}
\end{subfigure}
\hspace{3mm}
\begin{subfigure}[b]{0.47\linewidth}
\centering
\includegraphics[width=1.0\textwidth]{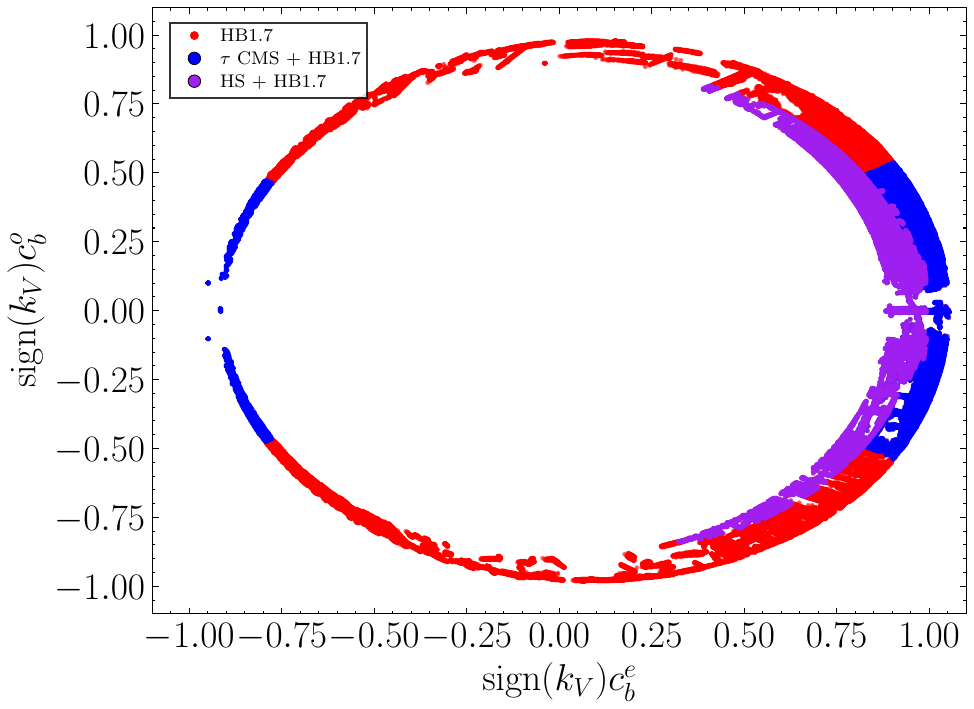}
\end{subfigure}
\caption{Same as in Fig.~\ref{fig:MLpoints_17_tau}, but for points which include the charm diagrams in the eEDM prediction.
}
\label{fig:MLpoints_charm_t2h2}
\end{figure}

Of course, although the charm contributions spoil the cancellation
of points obtained without it, new points can be found which, when including the charm contribution, produce new precise cancellations, in line with the experimental bound. Thus, we have performed a new simulation including both kite and charm contributions,
shown in Fig.~\ref{fig:MLpoints_charm_t2h2}. In the left panel, we show the results in the above defined $M_Z$-masses scheme and with the charm-quark mass chosen as $\overline{m}_c(\overline{m}_c)=1.2$~GeV. The right panel displays results obtained in the pole-masses scheme with the on-shell charm quark mass chosen as $m_c=1.51$~GeV\,\cite{Denner:2047636}. Comparing this figure (which includes the charm diagrams)
with Fig.~\ref{fig:MLpoints_17_tau} (which does not),
we see that all the features discussed previously remain.
In conclusion,
including the charm contributions alters the set of valid points obtained,
but it does not preclude compliance with the current eEDM experimental bounds,
nor does it alter the qualitative $c_b^o$ features.

\section{Revisiting other cases of the C2HDM}
\label{sec:others}
Having established the importance of both kite and charm contributions to the eEDM, we revisit all scenarios listed in Table~\ref{tab:ref} using ML techniques (Strategy~2). In the Type-I model, the CP-odd couplings of all fermions are universal and are therefore strongly constrained by the top-quark sector, forcing them to remain very close to their SM values for all choices of mass ordering. We further verify that no viable solutions with large $|c_b^o|$ components exist in the Type-II and Flipped models with $h_3 = h_{125}$, in agreement with the conclusions of Ref.~\cite{Fontes:2017zfn}.

\begin{figure}[h!]\centering
\begin{subfigure}[b]{0.47\linewidth}
\centering
\includegraphics[width=1.0\textwidth]{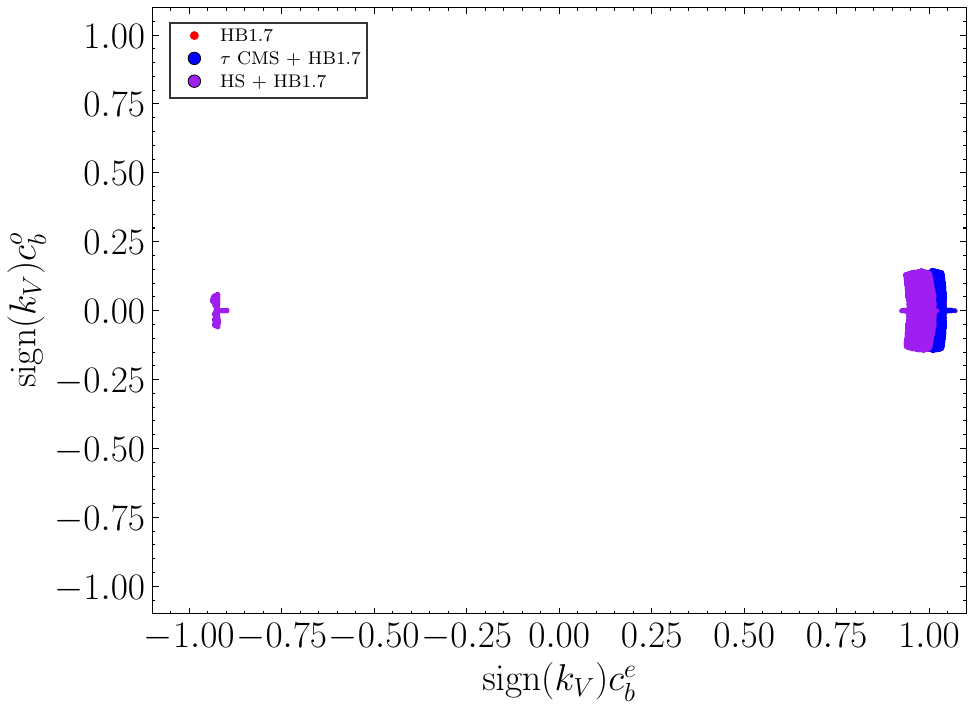}
\end{subfigure}
\hspace{3mm}
\begin{subfigure}[b]{0.47\linewidth}
\centering
\includegraphics[width=1.0\textwidth]{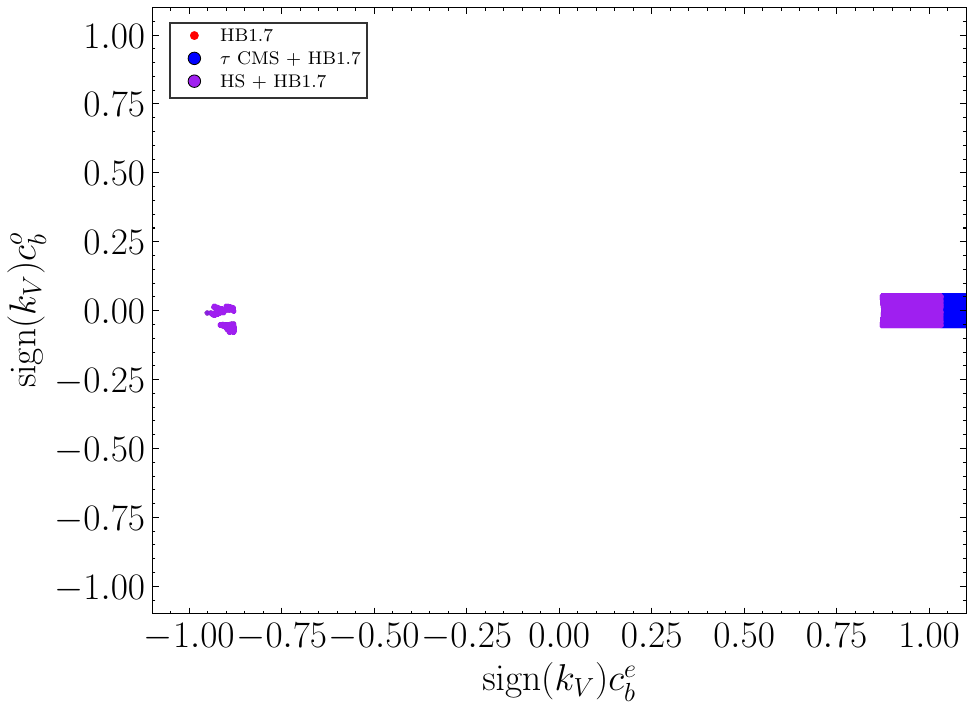}
\end{subfigure}
\caption{Points on the plane $\text{sign}(k_V)c_b^o \,\, \text{vs.} \,\, \text{sign}(k_V)c_b^e$ obtained with Strategy~2, in Type-II with $h_{125}=h_1$ (left) and Flipped with $h_{125}=h_1$ (right). In both panels, the eEDM is calculated with the pole-masses scheme. The color code is as in Fig.~\ref{fig:MLpoints_17_tau}.}
\label{fig:MLpoints_charm_h1}
\end{figure}

We also find no viable points with large $|c^o_b|$ for the mass choice $h_1=h_{125}$.
This is shown for Type-II and Flipped in Fig.~\ref{fig:MLpoints_charm_h1},
and the limitation is due to a
combination of searches for heavy resonances
\cite{ATLAS:2018sbw,ATLAS:2020tlo,ATLAS:2021fet}, which is checked by HT. Nevertheless, the recent Ref.~\cite{Lee:2025nuj} makes the point that there may be some large CP-violating signals in Type-II with $h_1=h_{125}$.

A few remarks are in order. First, looking at Fig.~13 in Ref.~\cite{Lee:2025nuj}, we agree that indeed there are no large $|c_b^o|$ couplings (for $h_{125}$). Second, even when $s_2=0$ (implying that $h_1$ is a pure scalar), Ref.~\cite{Barroso:2012wz} points out in their endnote 20 that there will still be CP violation in the heavier scalars,
proportional to $J_1 \propto (m_1^2 -m_2^2)(m_1^2 -m_3^2)(m_2^2 -m_3^2) \cos{(\beta -\alpha_1)} \sin^2{(\beta -\alpha_1)}
\cos{(\alpha_3)} \sin{(\alpha_3)}$. This source of CP violation only vanishes if the sine or cosine of $\beta - \alpha_1$ or of $\alpha_3$ vanish, or if the heavier two states are degenerate. That is, it vanishes in the exact alignment limit, where $h_1$ coincides exactly with the SM predictions, as re-emphasized in Ref.~\cite{Grzadkowski:2014ada}. What Ref.~\cite{Lee:2025nuj} brings to the table is that, as one goes slightly away from the exact alignment limit,  significant  CP-violating signals are possible in some observables, other than $c_b^o$.

\begin{figure}[h!]
\centering
\begin{subfigure}[b]{0.47\linewidth}
\centering
\includegraphics[width=1.0\textwidth]{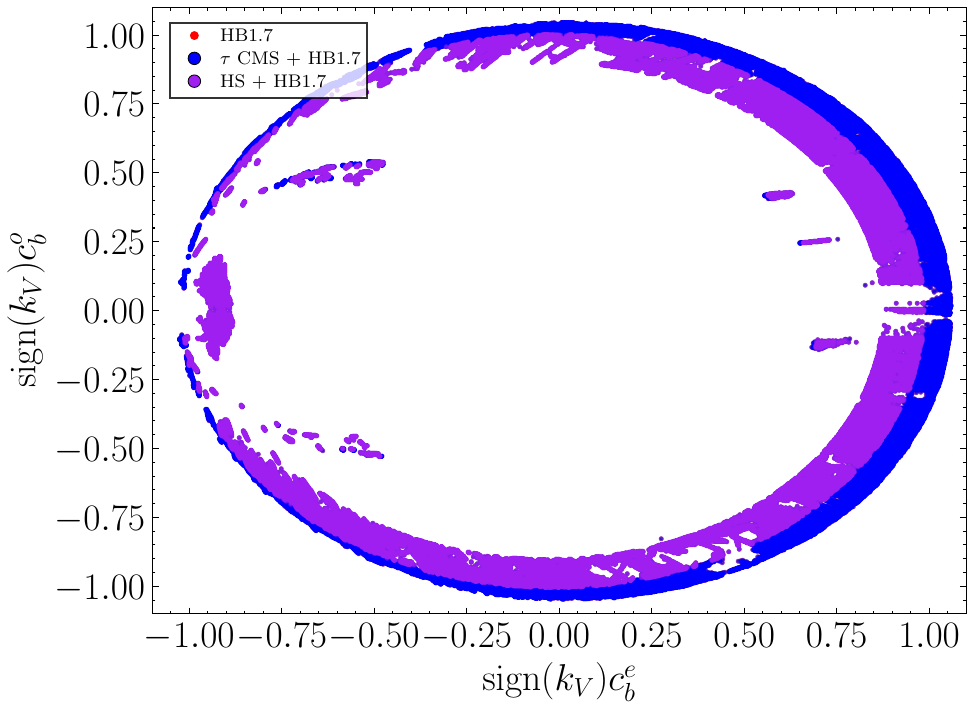}
\end{subfigure}
\hspace{3mm}
\begin{subfigure}[b]{0.47\linewidth}
\centering
\includegraphics[width=1.0\textwidth]{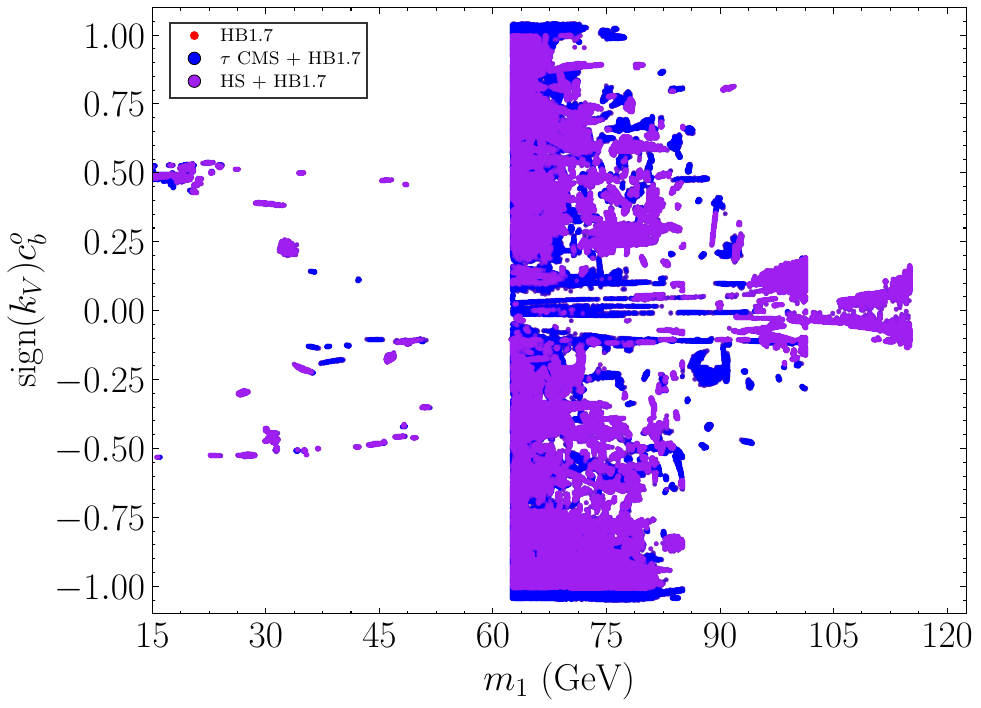}
\end{subfigure}
\caption{Flipped C2HDM with $h_2=h_{125}$, with the eEDM calculated with the $M_Z$-masses scheme, including both kite and charm contributions. The color code is as in Fig.~\ref{fig:MLpoints_17_tau}. Left: plane $\text{sign}(k_V)c_b^o \,\, \text{vs.} \,\, \text{sign}(k_V)c_b^e$. Right: plane $\text{sign}(k_V) c_b^o$ vs. $m_1$.}
\label{fig:t4h2_charm_mz}
\end{figure}

We now turn to the Flipped model, with $h_2=h_{125}$.
Our results are presented in Fig.~\ref{fig:t4h2_charm_mz}.
This case does allow for maximal $|c^o_b|$ couplings consistent with all current experimental data.
What this means is that, in this case,
one can have $h_{125}$ coupling mostly as a scalar to the top quark,
while it couples as a pure pseudoscalar to the bottom quark.
Again, we notice the presence of the isolated regions (with lower radius) with $m_{1} \leq 60$ GeV. For this case,
the points have dominant diagrams of order $10^{-29}\, \textrm{e.cm}$, unlike the order $10^{-27}\, \textrm{e.cm}$ individual contributions occurring in the case Type-II with $h_2=h_{125}$.

\section{Conclusions}
\label{sec:conclusions}
We have carried out a comprehensive exploration of the parameter space of the C2HDM using recent ML techniques designed to efficiently probe complex, high-dimensional BSM scenarios, even in the absence of pre-existing training data. To this end, we employed an Evolutionary Strategy algorithm \cite{deSouza:2022uhk} to identify phenomenologically viable points, combined with an anomaly-detection–based Novelty Reward mechanism \cite{Romao:2024gjx}. This framework enables a robust and reliable exploration of the model’s parameter space and, consequently, of its potentially observable physical predictions.

We have concentrated on regions of the parameter space allowing for large components of $c^o$, the pseudoscalar coupling between the 125 GeV Higgs boson ($h_{125}$) and fermions. In particular, we were interested in the striking possibility, first pointed out in Ref.~\cite{Fontes:2015mea}, that $h_{125}$ couples to the top quark as a pure scalar, while it couples to the bottom quark as a pure pseudoscalar.
Models with CP violation in the scalar sector usually must contend with large contributions to the eEDM, which is very constrained by current experiments to lie below $4.1 \times 10^{-30}\, \textrm{e.cm}$ \cite{Roussy:2022cmp}. Recent articles \cite{Altmannshofer:2020shb,Altmannshofer:2025nsl}
have pointed out the importance of kite diagrams
to the theoretical issue of gauge independence of the final result,
and to the practical issue of taking into account all numerically relevant
contributions.
Re-analyzing the C2HDM with the aid of ML,
we confirmed that kite diagrams are paramount,
and we further demonstrated that, at the current level of experimental precision, charm-quark Barr-Zee contributions must also be taken into account.

In the process, we investigated the sensitivity of the eEDM calculation to the renormalization prescriptions for the fermions masses,
by comparing $\overline{\rm MS}$ running
masses at the scale~$M_Z$ with pole masses.
We found that, using the kite diagrams, the qualitative predictions for $c^o$ made with either choice remain the same (although for different points in the model's parameter space).
Moreover, the cancellations required to obey the current eEDM bound are still possible if the bound is improved by three orders of magnitude.
As a result,
such measurements will not constrain the  significant  $c^o$ possibilities still available.
We thus strongly encourage a continued effort to improve the direct probes for $c^o$ at LHC.
We should add that, in the C2HDM, it is not possible to accommodate
large values for $c^o$ involving the top quark.
But that possibility does exist in the C3HDM, as shown in Refs.~\cite{deSouza:2025bpl,Boto:2025ovp}.
Hence, improved direct measurements of the CP-odd $h_{125}$ to the top quark
are also highly desirable.

Performing a full analyses of all model types and $h_{125}$ assignments,
we reach the conclusions shown in Table~\ref{tab:oursfinal}.
\begin{table}[ht]
\centering
\begin{tabular}{|l|c|c|c|c|}\hline
Type   & I & II& LS& Flipped\\ \hline
$h_1=h_{125}$&$\times$ &$\times$ &$\tau$ & $\times$\\\hline
$h_2=h_{125}$ &$\times$ & $\underline{\tau}$  &$\tau$ & $\underline{\checkmark}$ \\\hline
$h_3=h_{125}$ &$\times$ &$\times$ &$\tau$ & $\times$\\\hline
\end{tabular}
\caption{Updated version of Table \ref{tab:ref} after the analysis described in this paper. Underlined entries mark changes w.r.t.~Ref.~\cite{Biekotter:2024ykp}.
}
\label{tab:oursfinal}
\end{table}
In Type-II with $h_2=h_{125}$,
$c^o_b = c^o_\tau$ is only limited by LHC's searches for
CP violation in tau decays \cite{CMS:2021sdq,ATLAS:2022akr}; the results are shown in Fig.~\ref{fig:MLpoints_charm_t2h2}.
In contrast,
in the Flipped model with $h_2=h_{125}$,
$c^o_b$ can be maximal ($c^e_b=0$); the results are shown in Fig.~\ref{fig:t4h2_charm_mz}.
In this second case,
the wrong-sign solution ($c^e_b=-1, c^o_b=0$)
is also still viable.

In summary,
we have performed a full analysis of the C2HDM utilizing ML to
access the full parameter space and observable predictions,
complying with the full two-loop eEDM calculation.
The cases still viable call for increased experimental efforts
to probe CP-odd $h_{125}$ couplings to quarks directly.

\vspace{1ex}

\section*{Acknowledgments}
\noindent
We thank Wolfgang Altmannshofer for discussions. D.F. thanks Joachim Brod, Gino Isidori, Kilian Möhling, Ulrich Nierste, Zachary Polonsky and Dominik Stöckinger for discussions, and the Particle Theory Group at the UZH for its hospitality and support. 
The work of R.B., D.F. and M.M. is supported by the Deutsche Forschungsgemeinschaft (DFG, German Research Foundation) under grant 396021762-TRR 257. K.E.~acknowledges financial support from the Avicenna-Studienwerk.
The work of J.C.R., J.P.S., M.G. and R.S. is supported in part by the Portuguese
Fundação para a Ciência e Tecnologia (FCT) through the PRR (Recovery and Resilience
Plan), within the scope of the investment "RE-C06-i06 - Science Plus
Capacity Building", measure "RE-C06-i06.m02 - Reinforcement of
financing for International Partnerships in Science,
Technology and Innovation of the PRR", under the projects with
references 2024.01362.CERN and 2024.03328.CERN;
the work is
also supported by FCT under Contracts UIDB/00777/2020, UIDP/00777/2020 and UID/00618/2025. M.G. is additionally supported by FCT with a PhD Grant (reference 2023.02783.BD). The FCT projects are partially funded through
POCTI (FEDER), COMPETE, QREN and the EU.

\appendix

\section{\texorpdfstring{$\chi^2$ exclusion}{chisq exclusion}\label{app:chisqexclusion}}

As explained at the end of Sec.~\ref{sec:constraints}, using the public version of the HS module to impose bounds on $\theta_{\tau}$ leads to a systematic exclusion of parameter points. In this appendix, we provide a detailed illustration of this issue and clarify its origin.
For definiteness and ease of presentation, we focus on the representative case of Type-II with $h_2 = h_{125}$. We stress, however, that the behavior discussed below is generic and persists in all other realizations of the C2HDM.

\begin{figure}[h!]\centering
\begin{subfigure}[b]{0.42\linewidth}
\centering
\includegraphics[width=1.0\textwidth]{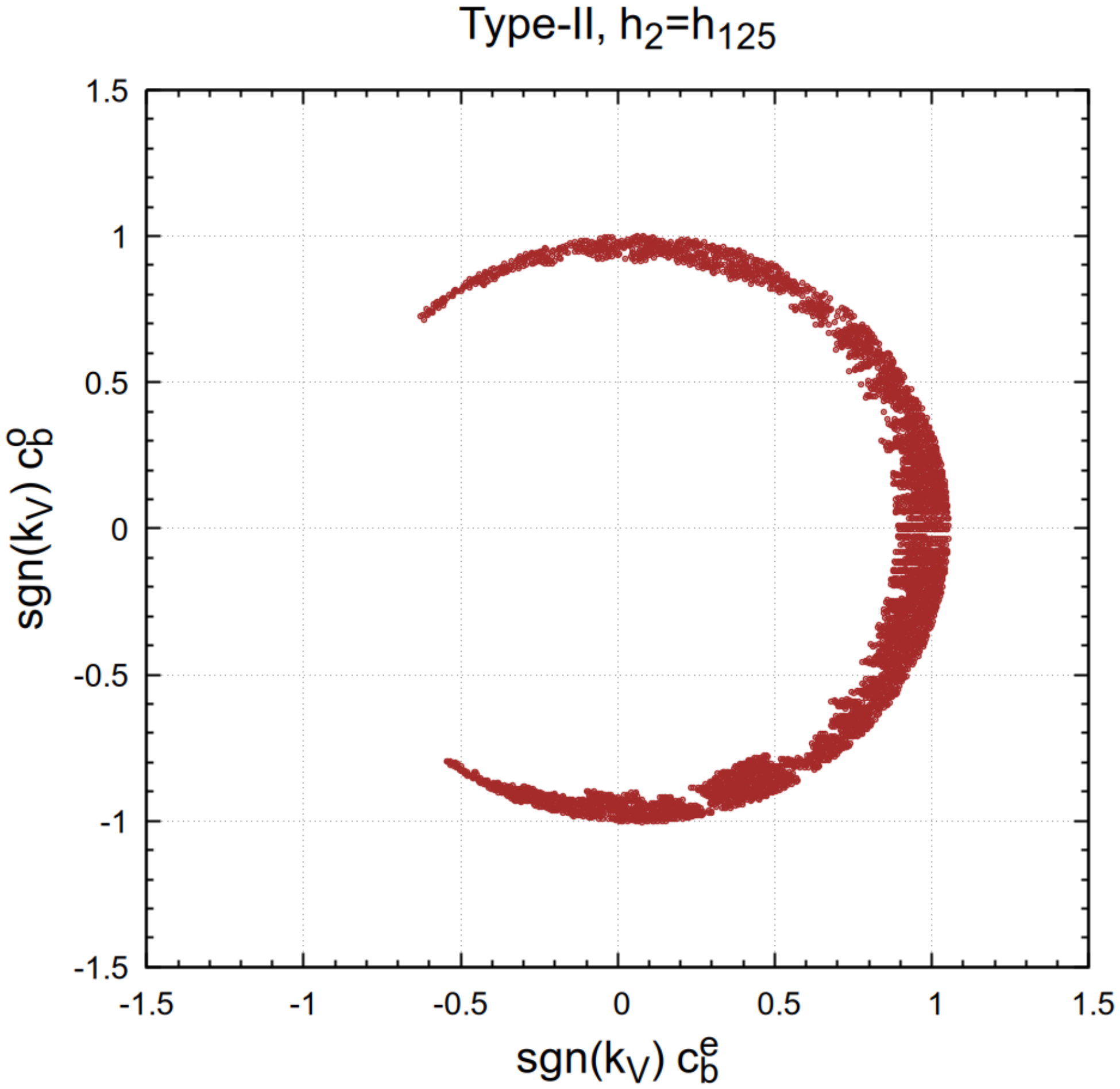}
\end{subfigure}
\hspace{1mm}
\begin{subfigure}[b]{0.52\linewidth}
\centering
\includegraphics[width=1.0\textwidth]{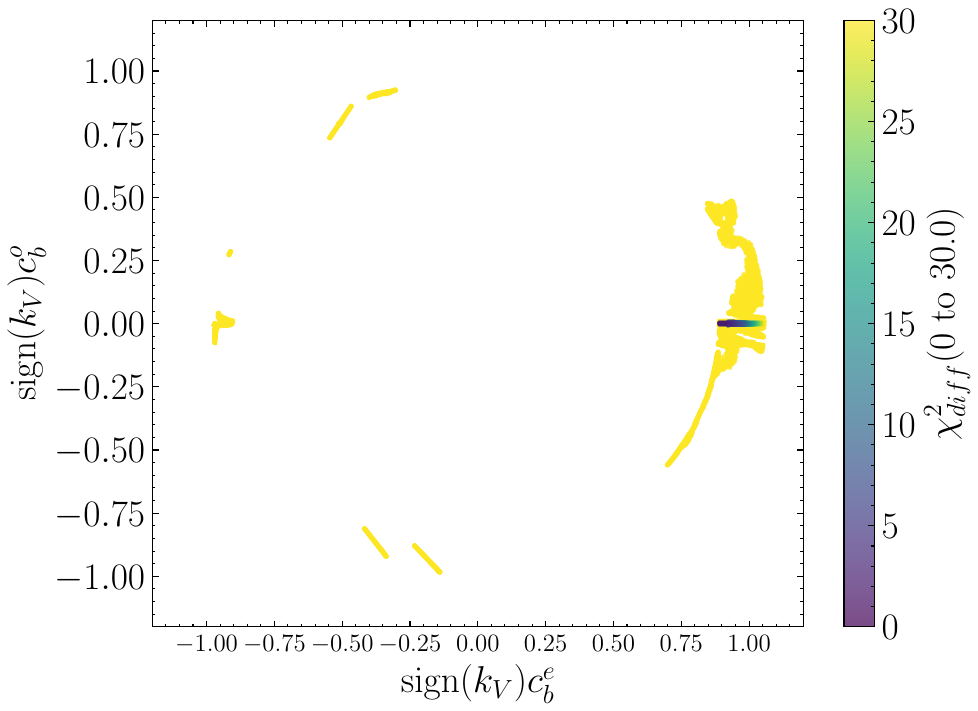}
\end{subfigure}
\caption{Points on the plane $\text{sign}(k_V)c_b^o \,\, \text{vs.} \,\, \text{sign}(k_V)c_b^e$ in the Type-II C2HDM with $h_{125}=h_2$ passing all constraints except those relative to $\theta_\tau$. Left: points obtained with Strategy~1. Right: points obtained with Strategy~2, with 10000 points per convergence and an unseeded run. On the right, all the (yellow) points that fail HS have $\chi^2_{\text{diff}}$ of order 150.
}
\label{fig:t2_h2_initial}
\end{figure}

The systematic exclusion can be seen by considering the left panel of Fig.~\ref{fig:t2_h2_initial}, which shows points that satisfy all constraints discussed in Sec.~\ref{sec:constraints} except those relative to $\theta_\tau$.
These points correspond exactly to the red points in the left panel of Fig.~\ref{fig:edmvary} and amount to a total of 4096 points.
The exclusion is verified by realizing that, once the  $\theta_\tau$ constraints are applied via HS,  less than 10 points survive. One might suspect that this dramatic reduction arises from limitations of the scanning procedure itself. After all, the left panel of Fig.~\ref{fig:t2_h2_initial} (like the corresponding panel of Fig.~\ref{fig:edmvary}) was obtained using Strategy 1, which relies on standard parameter scans. To test this hypothesis, we repeated the analysis using Strategy 2. The result is shown in the right panel of Fig.~\ref{fig:t2_h2_initial}, which displays the resulting values of $\chi^2_{\text{diff}}$, which are defined in Sec.~\ref{sec:constraints} and computed by HS. This plot demonstrates that the same pattern persists: only a very small number of points  yields acceptably low values of $\chi^2_{\text{diff}}$. Furthermore, all such points lie very close to the SM limit.

To better understand this issue, we isolate four points from Fig.~\ref{fig:t2_h2_initial}: two from the left panel (Points 1 and 3) and two from the right one (Points 2 and 4). The points are described in detail in Table \ref{table:chisqmeas}. Points 1 and 2 describe large pseudoscalar couplings, while Points 3 and 4 describe small ones. Only Point 3 passes HS, as it is the only one leading to a value of $\chi^2$ close to the SM, around 150. The remaining points describe a $\chi^2$ well above 150.
\begin{table}[h!]
\centering
\begin{tabular}{l | c | c | c | c }
\hline
Parameters & Point 1 & Point 2 & Point 3 & Point 4  \\
\hline
$m_{1}$ & 111.17800 & 117.61647 & 92.42560 & 109.20862  \\
$m_{2}$ & 125.09000 & 125.00000 & 125.09000 & 125.00000 \\
$m_{3}$ & 676.09279 & 676.12103 & 642.07726 & 647.28183  \\
$m_{H^\pm}$ & 675.95296 & 653.76571 & 641.68967 & 652.61301  \\
$\alpha_1$ & -0.12294140 & -0.09918623 & -0.20081799 & -0.17832599  \\
$\alpha_2$ & 0.42298603 & -1.00353030 & 0.00634782 & -0.00571511 \\
$\alpha_3$ & 0.10171668 & -0.18011110 & 0.00285759 & -0.00211328  \\
$\beta$ & 1.45177530 & 1.45820750 & 1.35522670 & 1.39790230 \\
\textrm{Re}($m_{12}^2$) & 10664.92700 & 37155.30600 & 1186.99800 & 1930.94000  \\
\hline
$c_{t}^o$ & -0.011073 & 0.01088342 & -0.00006257 & 0.00003691  \\
$c_{b}^o$ & -0.77427177 & 0.85132203 & -0.00130500 & 0.00121010 \\
$c_{\tau}^o$ & -0.77427177 & 0.85132203 & -0.00130500 & 0.00121010 \\
\hline 
$\mu_{ggF\times \gamma\gamma}$ & 0.8653  &  0.9369 &    0.9904 &  0.8637    \\
$\mu_{ggF\times ZZ}$&  0.9676 & 1.0771 & 1.1037   & 0.9572 \\   
$\mu_{ggF\times WW}$ & 0.9676   &  1.0771    &     1.1037  &       0.9572   \\   
$\mu_{ggF\times \tau\tau}$&  1.0371 &  1.0586  &     0.9599  &  1.0177   \\  
$\mu_{ggF\times bb}$&  1.0378  &     1.0595  &   0.9599 &   1.0177 \\
$\mu_{ggF\times Z\gamma}$ &  0.9298 &   1.0295 &   1.0610 &  0.9214    \\ 
\hline
$\chi^2$ & 333.89 & 351.95 & 154.55 & 224.20 \\
\hline
$\chi^2$ w/o Ref.~\cite{CMS:2021sdq} as a whole&   183.0112 & 185.3248& 153.5897 &  178.0807 \\ 
$\chi^2$ w/o $\mu_{ggH\times \tau\tau} = 0.59^{+0.28}_{-0.32}$  & 186.6076 & 195.5573  & 154.1852 & 178.4548 
\end{tabular}
\caption{Details of four points from Fig.~\ref{fig:t2_h2_initial}. In the public HS implementation, the SM reference value is $\chi^2_{\mathrm{SM}} = 152.54$. Removing the CMS:2021sd result \cite{CMS:2021sdq} in its entirety reduces this value to $\chi^2_{\mathrm{SM}} = 150.2$, while removing only the $\mu_{ggH\times\tau\tau}$ sub-measurement from CMS:2021sd \cite{CMS:2021sdq} yields $\chi^2_{\mathrm{SM}} = 150.8875$.
\label{table:chisqmeas}}
\end{table}

To investigate the origin of the values of $\chi^2$, we examine the individual contributions to $\chi^2$ of a certain point. This is done using the \texttt{chisqContributions(pred)} routine provided by HT, which allows for a comparison of all individual measurements.%
\footnote{See the HiggsTools documentation: \texorpdfstring{
\href{https://higgsbounds.gitlab.io/higgstools/HiggsSignalsAPI.html\#_CPPv4NK5Higgs7signals11Measurement18chisqContributionsERKN11predictions11PredictionsERK19ModificationFactors}%
  {HiggsSignals (chisqContributions)}%
}{HiggsSignals (chisqContributions)}}
As stated in the documentation, this procedure neglects correlations and multiplicative factors, implying that the sum of the individual contributions does not coincide with the $\chi^2$ obtained from the full combined measurement. Yet, this analysis reveals that Ref.~\cite{CMS:2021sdq} (corresponding to measurement number~13 in the HS public code) yields an anomalously large contribution to $\chi^2$. This is confirmed by explicitly removing this measurement from the calculation, as shown in the second-to-last line of Table~\ref{table:chisqmeas}.

\begin{figure}[h!]\centering
\begin{subfigure}[b]{0.46\linewidth}
\centering
\includegraphics[width=1.0\textwidth]{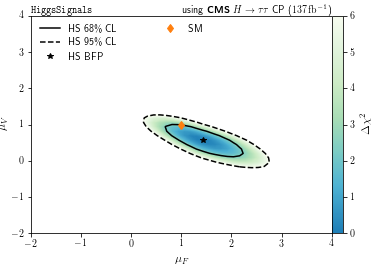}
\end{subfigure}
\hspace{3mm}
\begin{subfigure}[b]{0.49\linewidth}
\centering
\includegraphics[width=1.00\textwidth]{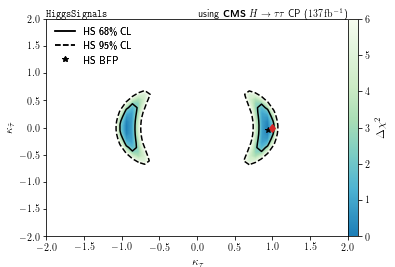}
\end{subfigure}
\caption{Figures taken from the public HS code, describing measurement number 13, relative to Ref.~\cite{CMS:2021sdq}. Left: reported value of $\mu_{ggH}^{\tau\tau} = 0.59^{+0.28}_{-0.32}$. 
Right: plane $\text{sign}(k_V)c_{\tau}^o \,\, \text{vs.} \,\, \text{sign}(k_V)c_{\tau}^e$. }
\label{fig:2021sdq}
\end{figure}

What exactly is the issue with Ref.~\cite{CMS:2021sdq}? 
This reference is implemented in HS as a multiplicity of submeasurements. Two of them are $\mu_{ggH}^{\tau\tau}$ and $\text{sign}(k_V)c_{\tau}^o \,\, \text{vs.} \,\, \text{sign}(k_V)c_{\tau}^e$, whose implementation in HS is shown in the left and right panels of Fig.~\ref{fig:2021sdq}, respectively. 
While the latter yields a physically sensible behavior --- crucial for the analysis presented in this work (see the discussion in Sec.~\ref{sec:Strategy2}) --- the former leads to an unexpected outcome: the SM point is excluded at the $1\sigma$ level. Importantly, this behavior originates from the experimental result reported in Ref.~\cite{CMS:2021sdq} itself and is not an artifact of its implementation within HS.
This observation can be placed in context by comparing it with the corresponding ATLAS measurements~\cite{ATLAS:2022vkf}, which yield $\mu_{ggF+bbH}^{\tau\tau} = 0.89788^{+0.29101}_{-0.25638}$ and $\mu_{ggF+bbH}^{bb} = 0.98031^{+0.37558}_{-0.36282}$. These results are consistent with the SM expectation and do not exhibit a similar tension.

\begin{figure}[h!]\centering
\begin{subfigure}[b]{0.46\linewidth}
\centering
\includegraphics[width=1.0\textwidth]{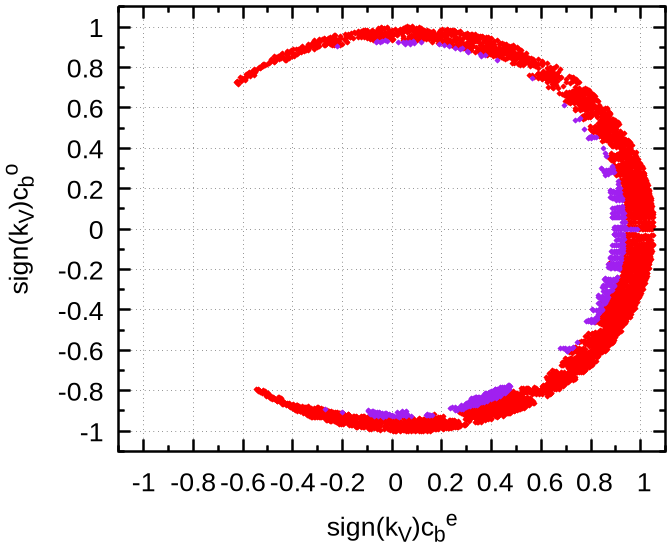}
\end{subfigure}
\hspace{7mm}
\begin{subfigure}[b]{0.46\linewidth}
\centering
\includegraphics[width=1.0\textwidth]{Figures/coups.png}
\end{subfigure}
\caption{Points on the plane $\text{sign}(k_V)c_b^o \,\, \text{vs.} \,\, \text{sign}(k_V)c_b^e$ in the Type-II C2HDM with $h_{125}=h_2$. Red points pass all constraints except those relative to $\theta_\tau$, violet points add these constraints using a modified HS: with Ref.~\cite{CMS:2021sdq} removed as a whole (left), or with only the $\mu_{ggH}^{\tau\tau}$ measurement removed (right). The right panel is identical to the left one of Fig.~\ref{fig:edmvary}.
}
\label{fig:mc_t2_h2_new}
\end{figure}

For this reason, we exclude the specific submeasurement of Ref.~\cite{CMS:2021sdq}'s $\mu_{ggH}^{\tau\tau}$ from the $\chi^2$ evaluation in HS.
In practice, this is achieved by commenting out the $\mu_{ggH}^{\tau\tau}$ entry in the corresponding HS dataset file
\url{https://gitlab.com/higgsbounds/hsdataset/-/blob/main/h125/tautau_CP_LHC13_CMS_137.json?ref_Type=heads}.
With this single submeasurement removed, the SM reference value is updated to
$\chi^2_{\mathrm{SM}} = 150.2$, compared to the original value
$\chi^2_{\mathrm{SM}} = 152.54$.
Reapplying HS to the parameter points shown in Fig.~\ref{fig:t2_h2_initial} then yields the results displayed in Fig.~\ref{fig:mc_t2_h2_new}, where 400 points out of the original 4096 are now found to be compatible with the HS constraints.

\bibliographystyle{JHEP}
\bibliography{ref}

\end{document}